\documentclass[prd,superscriptaddress,twocolumn,nofootinbib,showpacs,preprintnumbers,floatfix]{revtex4}

\usepackage{mathrsfs}
\usepackage{amsfonts,amssymb,amsmath}
\usepackage{graphicx,epsfig}
\usepackage{multirow}

\begin{document}

\title{Unification of Dynamical Determination and \\ Bare-Minimal Phenomenological Constraints in No-Scale $\boldsymbol{\mathcal{F}}$-$\boldsymbol{SU(5)}$}

\author{Tianjun Li}

\affiliation{State Key Laboratory of Theoretical Physics, Institute of Theoretical Physics,
Chinese Academy of Sciences, Beijing 100190, P. R. China }

\affiliation{George P. and Cynthia W. Mitchell Institute for Fundamental Physics and Astronomy,
Texas A$\&$M University, College Station, TX 77843, USA }

\author{James A. Maxin}

\affiliation{George P. and Cynthia W. Mitchell Institute for Fundamental Physics and Astronomy,
Texas A$\&$M University, College Station, TX 77843, USA }

\author{Dimitri V. Nanopoulos}

\affiliation{George P. and Cynthia W. Mitchell Institute for Fundamental Physics and Astronomy,
Texas A$\&$M University, College Station, TX 77843, USA }

\affiliation{Astroparticle Physics Group, Houston Advanced Research Center (HARC),
Mitchell Campus, Woodlands, TX 77381, USA}

\affiliation{Academy of Athens, Division of Natural Sciences,
28 Panepistimiou Avenue, Athens 10679, Greece }

\author{Joel W. Walker}

\affiliation{Department of Physics, Sam Houston State University,
Huntsville, TX 77341, USA }

%%%%%%%%%%%%%%%%%%%%%%%%%%%%%%%%%%%%%%%%%%%%%%%%%%%%%%%%%%%%%%%%%%%%%%%%%%%%

\begin{abstract}
We revisit the construction of the viable parameter space of No-Scale
$\cal{F}$-$SU(5)$, a model built on the $\cal{F}$-lipped $SU(5) \times U(1)_{\rm X}$
gauge group, supplemented by a pair of $\cal{F}$-theory derived vector-like multiplets at the TeV scale,
and the dynamically established boundary conditions of No-Scale Supergravity.  Employing an updated numerical
algorithm and a substantially upgraded computational engine,
we significantly enhance the scope, detail and accuracy of our prior study.  We sequentially apply
a set of ``bare-minimal'' phenomenological constraints, consisting of
i) the dynamically established boundary conditions of No-Scale Supergravity,
ii) consistent radiative electroweak symmetry breaking, iii) precision LEP constraints on
the light supersymmetric mass content, iv) the world average top-quark mass, and
v) a light neutralino satisfying the 7-year WMAP cold dark matter relic density measurement. 
The overlap of the viable parameter space with key rare-process limits on 
the branching ratio for $b \to s \gamma$ and the muon anomalous magnetic moment is
identified as the ``golden strip'' of $\cal{F}$-$SU(5)$.  A cross check for top-down theoretical
consistency is provided by application of the ``Super No-Scale'' condition, which dynamically selects
a pair of undetermined model parameters in a manner that is virtually identical to the corresponding
phenomenological (driven primarily by the relic density) selection.
The predicted vector-like particles are candidates for production at the future LHC,
which is furthermore sensitive to a distinctive signal of ultra-high multiplicity hadronic jets. 
The lightest CP-even Higgs boson mass is predicted to be $120^{+3.5}_{-1}$~GeV, with an additional
3--4~GeV upward shift possible from radiative loops in the vector-like multiplets.  The predominantly
bino flavored lightest neutralino is suitable for direct detection by the Xenon collaboration.
\end{abstract}


\preprint{ACT-06-11, MIFPA-11-16}

\maketitle

%%%%%%%%%%%%%%%%%%%%%%%%%%%%%%%%%%%%%%%%%%%%%%%%%%%%%%%%%%%%%%%%%%%%%%%%%%%%

\section{Introduction\label{sct:intro}}

The driving aim of theoretical physics is to achieve maximal efficiency in the correlation of observations.
This entails the unification of apparently distinct forces under a master symmetry group, and the
successful reinterpretation of experimentally constrained parameters and finely tuned scales
as dynamically evolved consequences of the underlying equations of motion.
The great challenge of string phenomenology is the construction of realistic models featuring clear predictions,
preferably uniquely indicative of their stringy origin, which can be definitively tested at the Large Hadron Collider
(LHC), future International Linear Collider (ILC), or other current and near term high energy experiments such as those
focused on the direct detection of dark matter or proton decay.  The LHC at CERN has been accumulating 
data from ${\sqrt s}=7$ TeV proton-proton collisions since March 2010, and has reportedly delivered 
an integrated luminosity of $5~{\rm fb}^{-1}$ to each major detector at the close of 2011. It is expected that this
number may quadruple to $20~{\rm fb}^{-1}$ by the end of 2012, making the search for such models
an issue of immediate relevance and opportunity. 

In a series of recent and contemporaneous publications~\cite{
Li:2010ws, Li:2010mi,Li:2010uu,Li:2011dw, Li:2011hr, Maxin:2011hy,
Li:2011in,Li:2011gh,Li:2011rp,Li:2011fu,Li:2011xg,Li:2011ex,Li:2011av,Li:2011ab},
we have studied in some substantial detail a promising model by the name of No-Scale $\cal{F}$-$SU(5)$, which 
is constructed from the merger of the ${\cal F}$-lipped $SU(5)$ Grand Unified Theory
(GUT)~\cite{Barr:1981qv,Derendinger:1983aj,Antoniadis:1987dx},
two pairs of hypothetical TeV scale vector-like supersymmetric (SUSY) multiplets with origins in
${\cal F}$-theory~\cite{Jiang:2006hf,Jiang:2009zza,Jiang:2009za,Li:2010dp,Li:2010rz},
and the dynamically established boundary conditions of No-Scale
Supergravity~\cite{Cremmer:1983bf,Ellis:1983sf, Ellis:1983ei, Ellis:1984bm, Lahanas:1986uc}.
Having demonstrated in turn the model's broad phenomenological consistency, profound predictive capacity,
singularly distinctive experimental signature and imminent testability, we begin now to retrace our initial
steps, seeking to revise, expand, and reflect -- with the wider view afforded by some distance -- upon that
fledgling analysis.  The intervening season of study has incubated a fresh flowering of refinements in our
numerical technique and likewise also in our conceptual grip on the model's internal dynamics, rendering a return
to these results both vital and timely.  We find in several cases a simple validation of prior work, and in certain
others, that our preliminary conclusions require a not insubstantial modification; however, despite certain numerical
parameter reassignments, we find the coherence of the underlying construction to be in all regards undiminished.

The central focus of the present work will be a phenomenologically driven bottom-up survey of the parameter
space of No-Scale $\cal{F}$-$SU(5)$, which dramatically expands the scope of our earlier numerical scans
by leveraging significantly upgraded computational resources, and substantial coding refinements.  We thus first establish
the full exterior borders of the ``bare-minimally'' allowed $\cal{F}$-$SU(5)$ model, distinguished by consistency
with a subset of experimental and theoretical constraints of the utmost stability and criticality.  A ``golden strip''
within this larger space is isolated by the further adherence to limits on rare processes, namely the SUSY contributions
to flavor-changing neutral currents and the anomalous magnetic moment of the muon.  This effort is cross-referenced
against the theoretical top-down perspective, by selective application of the ``Super No-Scale'' condition~\cite{Li:2010uu,Li:2011dw}
to dynamically isolate preferred parameterizations; interestingly, we find that the procedural modifications presented in this work are
precisely such to shift the phenomenologically favored space, and particularly the ratio $\tan\beta$ of up- and down-like
Higgs vacuum expectation values (VEVs), into an unprecedented precision of convergence with the dynamical determination.

The two key improvements in our numerical treatment are a fix at the sub-integral level to the resolution of
the vector-like multiplet $\beta$-function coefficients, and a more direct evaluation
of the key $B_\mu = 0$ boundary condition at the high scale itself, rather than as a matching condition
on the magnitude of $B_\mu$ at the point of consistent electroweak symmetry breaking (EWSB).
We are also now technologically more capable, having developed a robust procedural automation 
and an enhanced systematic integration of the requisite computational phases, including our proprietary
modifications to the {\tt MicrOMEGAs 2.1}~\cite{Belanger:2008sj} and {\tt SuSpect 2.34}~\cite{Djouadi:2002ze}
code bases, plus all internally essential data post-processing.
The construction of a custom wrapper written in the Message Passing Interface (MPI) protocol has allowed
us to efficiently scale up the numerics to run within a high performance parallel computing environment.
In conjunction, these upgrades have facilitated new scans of unprecedented scope and detail,
providing a previously unavailable wide angle view of the No-Scale $\cal{F}$-$SU(5)$ parameter space.
We emphasize that this enlargement of our parameterization under the perspective of the bare-minimal constraints
is not a refutation of the more narrowly focused application of constraints from earlier work~\cite{Li:2010ws,Li:2010mi},
but rather a complementary approach, based upon a distinct set of opening philosophical assumptions.

%%%%%%%%%%%%%%%%%%%%%%%%%%%%%%%%%%%%%%%%%%%%%%%%%%%%%%%%%%%%%%%%

\section{The No-Scale $\boldsymbol{\cal{F}}$-$\boldsymbol{SU(5)}$ Model}

The No-Scale $\cal{F}$-$SU(5)$ construction inherits all of the most beneficial phenomenology~\cite{Nanopoulos:2002qk}
of flipped $SU(5)$~\cite{Barr:1981qv,Derendinger:1983aj,Antoniadis:1987dx},
including fundamental GUT scale Higgs representations (not adjoints), natural doublet-triplet
splitting, suppression of dimension-five proton decay and a two-step see-saw mechanism
for neutrino masses, as well as all of the most beneficial theoretical motivation of No-Scale
Supergravity~\cite{Cremmer:1983bf,Ellis:1983sf, Ellis:1983ei, Ellis:1984bm, Lahanas:1986uc},
including a deep connection to the string theory infrared limit
(via compactification of the weakly coupled heterotic theory~\cite{Witten:1985xb} or 
M-theory on $S^1/Z_2$ at the leading order~\cite{Li:1997sk}),
the natural incorporation of general coordinate invariance (general relativity),
quantum stabilization of the electroweak (EW) gauge hierarchy by supersymmetry (SUSY),
a natural cold dark-matter (CDM) candidate in the form of the
lightest supersymmetric particle (LSP) ~\cite{Ellis:1983ew,Goldberg:1983nd},
a mechanism for SUSY breaking which preserves a vanishing cosmological constant at the tree level
(facilitating the observed longevity and cosmological flatness of our Universe~\cite{Cremmer:1983bf}),
natural suppression of CP violation and flavor-changing neutral currents, dynamic stabilization
of the compactified spacetime by minimization of the loop-corrected scalar potential and a dramatic
reduction in parameterization freedom.

Written in full, the gauge group of flipped $SU(5)$ is $SU(5)\times U(1)_{X}$, which can be embedded into $SO(10)$.
The generator $U(1)_{Y'}$ is defined for fundamental five-plets as $-1/3$ for the triplet members, and $+1/2$ for the doublet.
The hypercharge is given by $Q_{Y}=( Q_{X}-Q_{Y'})/5$.  There are three families of Standard Model (SM) fermions,
whose quantum numbers under the $SU(5)\times U(1)_{X}$ gauge group are
\begin{equation}
F_i={\mathbf{(10, 1)}} \quad;\quad {\bar f}_i={\mathbf{(\bar 5, -3)}} \quad;\quad {\bar l}_i={\mathbf{(1, 5)}},
\label{eq:smfermions}
\end{equation}
where $i=1, 2, 3$.  There is a pair of ten-plet Higgs for breaking the GUT symmetry, and a pair
of five-plet Higgs for electroweak symmetry breaking (EWSB).
\begin{eqnarray}
& H={\mathbf{(10, 1)}}\quad;\quad~{\overline{H}}={\mathbf{({\overline{10}}, -1)}} & \nonumber \\
& h={\mathbf{(5, -2)}}\quad;\quad~{\overline h}={\mathbf{({\bar {5}}, 2)}} &
\label{eq:Higgs}
\end{eqnarray}
Since we do not observe mass degenerate superpartners for the known SM fields, SUSY must itself be broken around the TeV scale.
In the minimal supergravities (mSUGRA), this occurs first in a hidden sector, and the secondary propagation by gravitational interactions
into the observable sector is parameterized by universal SUSY-breaking ``soft terms'' which include the gaugino mass $M_{1/2}$, scalar mass
$M_0$ and the trilinear coupling $A$.  The ratio of the low energy Higgs VEVs $\tan \beta$, and the sign of
the SUSY-preserving Higgs bilinear mass term $\mu$ are also undetermined, while the magnitude of the $\mu$ term and its bilinear soft term $B_{\mu}$
are determined by the $Z$-boson mass $M_{\rm Z}$ and $\tan \beta$ after EWSB.  In the simplest No-Scale scenario,
$M_0$=A=$B_{\mu}$=0 at the unification boundary, while the complete collection of low energy SUSY breaking soft-terms evolve down 
from a single non-zero parameter $M_{1/2}$.  Consequently, the particle spectrum will be proportional to $M_{1/2}$ at leading order,
rendering the bulk ``internal'' physical properties invariant under an overall rescaling.
The rescaling symmetry can likewise be broken to a certain degree by the vector-like
mass parameter $M_{\rm V}$, although this effect is weak.

The matching condition between the low-energy value of $B_\mu$ that is demanded by EWSB and
the high-energy $B_\mu = 0$ boundary is notoriously difficult to reconcile under the
renormalization group equation (RGE) running in a phenomenologically consistent manner.
The present solution relies on modifications to the $\beta$-function coefficients that are generated by the inclusion
of the extra vector-like multiplets, which may actively participate in radiative loops above their characteristic mass threshold $M_{\rm V}$. 
Naturalness in view of the gauge hierarchy and $\mu$ problems suggests that the mass $M_{\rm V}$ should be of the TeV order.
Avoiding a Landau pole for the strong coupling constant restricts the set of vector-like multiplets which may be
given a mass in this range to only two constructions with flipped charge assignments, which have been explicitly realized
in the $F$-theory model building context~\cite{Jiang:2006hf,Jiang:2009zza, Jiang:2009za}.  We adopt the following two
multiplets, along with their conjugates, where $XQ$, $XD^c$, $XE^c$ and $XN^c$ carry the same quantum
numbers as the quark doublet, right-handed down-type quark, charged lepton and neutrino, respectively.
\begin{equation}
{XF}_{\mathbf{(10,1)}} \equiv (XQ,XD^c,XN^c) \quad;\quad {\overline{Xl}}_{\mathbf{(1, 5)}} \equiv XE^c \,
\label{eq:flippons}
\end{equation}
Alternatively, the pair of $SU(5)$ singlets $(Xl,\overline{Xl})$ may be discarded, but phenomenological consistency then
requires the substantial application of unspecified GUT thresholds.  In either case, the (formerly negative) one-loop $\beta$-function
coefficient of the strong coupling $\alpha_3$ becomes precisely zero, flattening the RGE running, and generating a wide
gap between the large $\alpha_{32} \simeq \alpha_3(M_{\rm Z}) \simeq 0.11$ and the much smaller $\alpha_{\rm X}$ at the scale $M_{32}$ of the intermediate
flipped $SU(5)$ unification of the $SU(3) \times SU(2)_{\rm L}$ subgroup.  This facilitates a very significant secondary running phase
up to the final $SU(5) \times U(1)_{\rm X}$ unification scale $M_{\cal{F}}$~\cite{Li:2010dp}, which may be elevated by 2-3 orders of magnitude
into adjacency with the Planck mass, where the $B_\mu = 0$ boundary condition fits like hand to glove~\cite{Ellis:2001kg,Schmaltz:2000gy,Ellis:2010jb,Li:2010ws}.  
This natural resolution of the ``little hierarchy'' problem corresponds also to true string-scale gauge coupling unification in
the free fermionic string models~\cite{Jiang:2006hf,Lopez:1992kg} or the decoupling scenario in F-theory models~\cite{Jiang:2009zza,Jiang:2009za},
and also helps to address the monopole problem via hybrid inflation.

The modifications to the $\beta$-function coefficients from introduction of the vector-like multiplets have a
parallel effect on the RGEs of the gauginos.  In particular, the color-charged gaugino mass $M_{\rm 3}$ likewise runs down flat from the
high energy boundary, obeying the relation $M_3/M_{1/2} \simeq \alpha_3(M_{\rm Z})/\alpha_3(M_{32}) \simeq \mathcal{O}\,(1)$,
which precipitates a conspicuously light gluino mass assignment.
The $SU(2)_{\rm L}$ and hypercharge $U(1)_{\rm Y}$ associated gaugino masses are by
contrast driven downward from the $M_{1/2}$ boundary value by roughly the ratio of their corresponding gauge couplings
$(\alpha_2,\alpha_{\rm Y})$ to the strong coupling $\alpha_{\rm s}$.
The large mass splitting expected from the heaviness of the top quark via its strong coupling to the Higgs (which is also key to generating an appreciable
radiative Higgs mass shift $\Delta~m_h^2$~\cite{Li:2011ab}) is responsible for a rather light stop squark $\widetilde{\rm t}_1$.
The distinctively predictive $M({\widetilde{t_1}}) < M({\widetilde{g}}) < M({\widetilde{q}})$ mass hierarchy of a light stop
and gluino, both much lighter than all other squarks, is stable across the full No-Scale $\cal{F}$-$SU(5)$ model space, but is
not precisely replicated in any phenomenologically favored constrained MSSM (CMSSM) constructions of which we are aware.

This spectrum generates a unique event topology starting from the pair production of heavy squarks
$\widetilde{q} \widetilde{\overline{q}}$, except for the light stop, in the initial hard scattering process,
with each squark likely to yield a quark-gluino pair $\widetilde{q} \rightarrow q \widetilde{g}$.  Each gluino may be expected
to produce events with a high multiplicity of virtual stops, via the (possibly off-shell) $\widetilde{g} \rightarrow \widetilde{\rm t}$
transition, which in turn may terminate into hard scattering products such as $\rightarrow W^{+}W^{-} b \overline{b} \widetilde{\chi}_1^{0}$
and $W^{-} b \overline{b} \tau^{+} \nu_{\tau} \widetilde{\chi}_1^{0}$, where the $W$ bosons will produce mostly hadronic jets and some leptons.
The model described may then consistently exhibit a net product of eight or more hard jets emergent from a single squark pair production event,
passing through a single intermediate gluino pair, resulting after fragmentation in a spectacular signal of ultra-high multiplicity final state jet events.
We remark also that the entirety of the viable $\cal{F}$-$SU(5)$ parameter space naturally features a
dominantly bino LSP, at a purity greater than 99.7\%, as is exceedingly suitable for direct detection, for example by
XENON100~\cite{Aprile:2011hi,Li:2011in}.  There exists no direct bino to wino mass mixing term.
This distinctive and desirable model characteristic is guaranteed by the relative heaviness of the Higgs bilinear
mass $\mu$, which in the present construction generically traces the universal gaugino mass $M_{1/2}$ at the boundary
scale $M_{\cal F}$, and subsequently transmutes under the RGEs to a somewhat larger value at the electroweak scale.

We reserve analysis of the prospective influence of Yukawa-coupled radiative loops in the vector-like
multiplets on the running of the renormalization group for future work.  These contributions are
expected to be rather small, entering at the second order for the evolution of the gauge couplings,
and at single loop for $\mu$ and $B_\mu$~\cite{Martin:1993zk}.  Uncertainty in the vector-like mass scale $M_{\rm V}$
(in contrast to the well known top quark mass, whose second-loop contributions are tallied) introduces
fluctuations in the mass-threshold of the leading single-loop contributions that may potentially envelop the amplitude
of the second order.  Likewise, a stringy no-scale supergravity construction will itself be subject to
corrections in the next order, and in particular, to modifications of the soft SUSY-breaking parameters
(such as $B_\mu$) at the high scale.  Since the leading intention of the present analysis is a determination
of the ``bare-minimal'' constraints on the viable model space of our theory, we judge that such a focus on
small local modifications to the scale or interplay of the $M_{1/2}, M_{\rm V}, \tan \beta$ and
$m_{\rm t}$ parameterization would run somewhat counter the established tone,
without substantively altering the globally established model perimeter.
An occasion for such refinements may exist after further testing by the LHC of the bulk lower order
model predictions, if the leading experimental indications at that time should remain positive.

%%%%%%%%%%%%%%%%%%%%%%%%%%%%%%%%%%%%%%%%%%%%%%%%%%%%%%%%%%%%%%%%

\section{The Bare-Minimal Constraints\label{sct:baremin}}

\begin{figure*}[htf]
        \centering
        \includegraphics[width=0.95\textwidth]{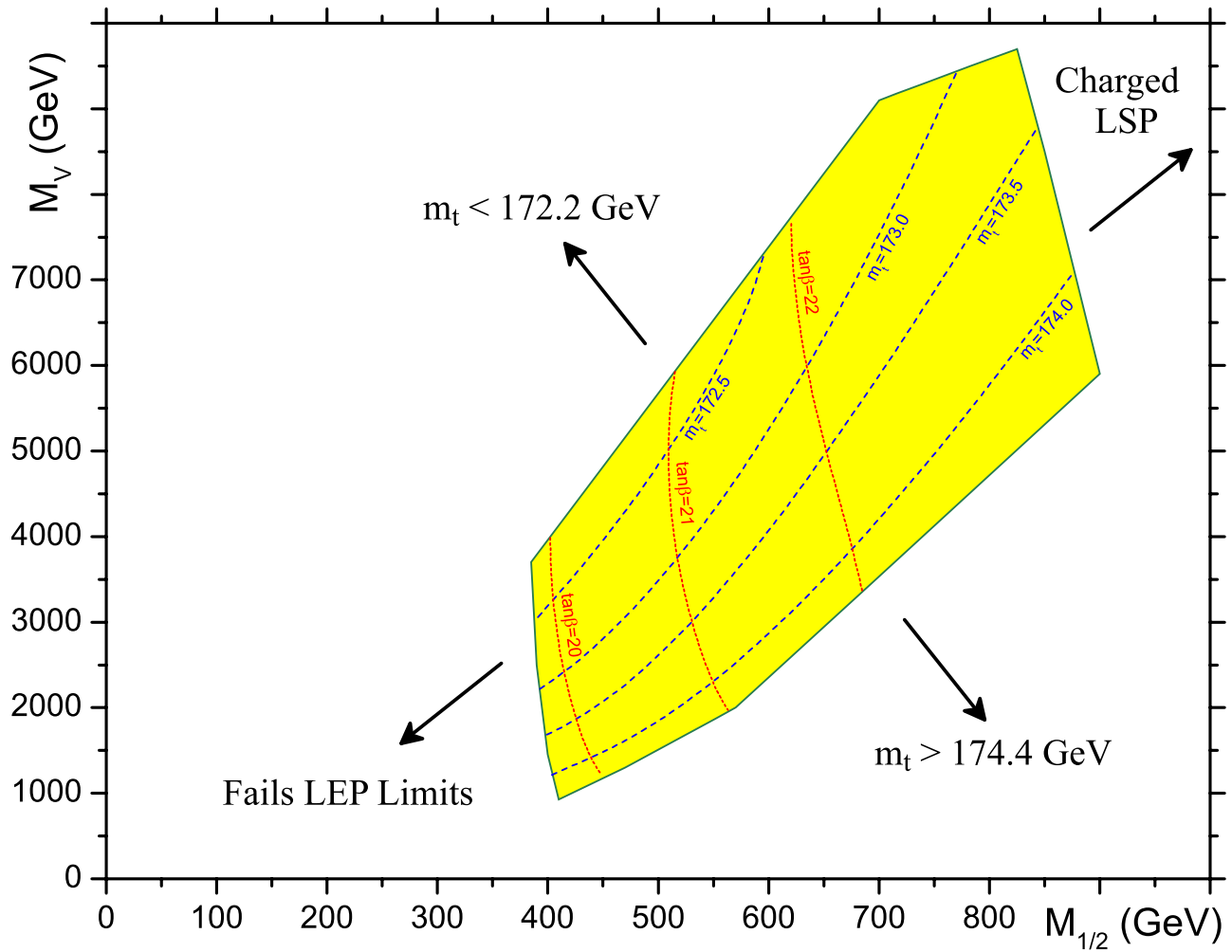}
        \caption{The surviving $(M_{1/2}, M_{\rm V}, m_{\rm t}, \tan \beta)$ parameter space is depicted
following application of the bare-minimal experimental constraints, with the exclusion regions noted.
The shaded region satisfies
i) the dynamically established high-scale boundary conditions $M_0=A=B_{\mu}=0$ of No-Scale Supergravity,
ii) consistent radiative electroweak symmetry breaking,
iii) precision LEP constraints on the lightest CP-even Higgs boson $m_{h}$ and other light SUSY chargino and neutralino mass content,
iv) the world average top-quark mass $172.2~{\rm GeV} \leq m_{\rm t} \leq 174.4~{\rm GeV}$, and
v) the 7-year WMAP limits 0.1088 $\leq \Omega_{\rm CDM} \leq$ 0.1158 with a single, neutral LSP as the CDM candidate.}
        \label{fig:Wedge_BareMinimal}
\end{figure*}

We adopt here certain distinctions in our phenomenological perspective relative to prior work.  Specifically, we
have chosen to impose the relevant empirical constraints hierarchically, emphasizing first those results which we perceive
to have been established in the broadest and most direct manner, and upon which there is the greatest consensus
with regards to basic stability of the experimental conclusion.  We will refer to this data subset, in conjunction with
the theoretically defining boundary conditions of the No-Scale models, as the ``bare-minimal'' constraints, and will define
and discuss each element of this set in sequence.  The surviving parameter space of our four scanning variables
$(M_{1/2}, M_{\rm V}, m_{\rm t}, \tan \beta)$ is depicted in Figure~\ref{fig:Wedge_BareMinimal}.

The leading criterion of our bare-minimal constraints represent the enforcement of the defining dynamic boundary conditions
of No-Scale Supergravity.  As suggested in the prior section, these include the vanishing at some high scale of the universal scalar
mass $M_0$ and the tri/bi-linear soft SUSY breaking couplings $A$ and $B_\mu$, related respectively to the SUSY preserving Yukawa
interaction and Higgs mixing mass term $\mu H_d H_u$.
Whereas $M_0 = A = 0$ may be imposed directly, the nullification of $B_\mu$ is rather more subtle, in that this parameter
is usually interpreted as one of two dependent outputs of the EWSB minimization procedure (the other being $\mu$ itself), and
we must then impose a consistency condition on the RGE evolved value of $B_\mu$ measured at the boundary scale.
This matching is notoriously difficult to reconcile while otherwise maintaining acceptable phenomenology, 
and for all the theoretical heft of the No-Scale construction, when this boundary
is applied at a traditional GUT scale, it simply does not work~\cite{Ellis:2001kg,Schmaltz:2000gy,Ellis:2010jb}.  The situation
may be dramatically improved by elevating the upper scale into closer proximity to the reduced Planck
mass~\cite{Ellis:2001kg,Schmaltz:2000gy,Ellis:2010jb,Li:2010ws}; a particularly natural and satisfactory assignment of the
No-Scale boundary scale may be made at the point $M_{\cal{F}}$ where the flipped $SU(5) \times U(1)_{\rm X}$ gauge group
realizes its final unification.  Even so, $B_\mu(M_{\cal{F}}) = 0$ remains a particularly strong condition, which dramatically
reduces the allowed parameter space.  It must also be emphasized, however, that the four scanning degrees of freedom can and will
conspire by intra-compensatory variation to define a large hyper-surface of acceptable solutions for the No-Scale boundary condition.
As has been our practice~\cite{Li:2010mi}, we allow an uncertainty of $\pm 1$~GeV on $B_\mu = 0$, consistent with the induced variation
from fluctuation of the strong coupling within its error bounds, and likewise with the expected scale of radiative EW corrections.

The second and third tiers of our bare-minimal constraints have to do with enforcing internal consistency of the
renormalization group and the broad phenomenology of the resulting SUSY spectrum.  We begin by rejecting any parameter
combinations which are incapable of spontaneously destabilizing the Higgs vacuum via the process of radiative EWSB,
as triggered by the condition $M^2_{H_u}+\mu^2 < 0$, where the rapid descent of the up-like Higgs mass-square $M^2_{H_u}$
is driven by the large Yukawa associated with the heaviness of the top quark.  Next, we perform a systematic consistency
check of precision LEP constraints on the lightest CP-even Higgs boson ($m_{h} \geq 114$ GeV~\cite{Barate:2003sz,Yao:2006px})
and other light SUSY chargino, stau, and neutralino mass content, as facilitated by the {\tt MicrOMEGAs 2.1}~\cite{Belanger:2008sj}
software package.  A lower bound on $M_{1/2}$ around 385~GeV and also an associated lower bound on $\tan \beta$ around 19.4
are established by the LEP constraints, as noted within Figure~\ref{fig:Wedge_BareMinimal}.

We treat the top quark mass as a scanning parameter, based on the substantial leverage it exerts over
the renormalization group equations (RGEs) via proportionality to the dominant Yukawa coupling, and the allowed
range of this parameter constitutes the fourth tier of the bare-minimal constraints.  We place the top
quark firmly within the conceptual category of experimental input in this work (see Ref.~\cite{Li:2010mi} for an alternate
possible point of view), and establish its permissible variation to be within the experimental bounds of
$m_{\rm t}$ = $173.3\pm 1.1$ GeV~\cite{:1900yx}.  The remaining three scanning parameters, consisting of $\tan \beta$,
the vector-like mass scale $M_{\rm V}$, and the universal gaugino boundary mass $M_{1/2}$ have no direct experimental bounds,
but we do select a sufficiently wide array of possibilities to ensure abutment against an imposed constraint of some other variety.
It must be emphasized that it is the combination of the $B_\mu = 0$ boundary and the relic density condition to be described subsequently
which associates a single point from the hidden $(m_{\rm t}$,$\tan \beta)$ plane with each visible point within
the $(M_{1/2},M_{\rm V})$ plane.  As demonstrated in Figure~\ref{fig:Wedge_BareMinimal}, the lower limit on $m_{\rm t}$ creates a
diagonal exclusion boundary at the upper left of the $(M_{1/2},M_{\rm V})$ plane, beyond which variation in the other parameters
is incapable of restoring consistency; the upper limit on $m_{\rm t}$ creates a corresponding exclusion at the lower right.

The fifth and final of the bare-minimal constraints regards the prediction of an appropriate single component source of
the WMAP7 observed cold dark matter relic density, including a strict barrier against a charged particle appearing as the LSP.
Specifically, we enforce the $0.1088 \leq \Omega_{\rm CDM} \leq 0.1158$~\cite{Komatsu:2010fb}, although one may make some distinction
between the upper and lower bounds here, insomuch as the lower bound may be relaxed in a multi-component dark matter scenario.
Our clear personal bias, however, is a neutralino dominated dark matter density, and the stable No-Scale $\cal{F}$-$SU(5)$ prediction of
a bino flavored LSP fits the bill rather nicely.  Because the spin-independent annihilation cross section is about $2\times 10^{-10}$ pb, 
it is an excellent candidate for near term direct detection by the Xenon collaboration~\cite{Aprile:2011hi}, which
has some realistic hopes of trumping the collider based search for signs of supersymmetry at the LHC and the Tevatron.
The CDM relic density target, along with the $B_\mu = 0$ boundary condition, each play the important role of removing
a degree of freedom from the parameterization, such that a fixed point in the $(M_{1/2},M_{\rm V})$ plane will be in monotonic
correspondence with a unique $(m_{\rm t},\tan\beta)$ pairing, modulo variation within some uncertainty.
We find rather generically that the stau $\tilde{\tau}$, SUSY partner to the tau $\tau$, will take over the LSP role if the ratio
$\tan \beta$ advances above a value of about $23$ as noted in Figure~\ref{fig:Wedge_BareMinimal}.
A strong correlation between $\tan \beta$ and $M_{\rm V}$ under the dynamics imposed by the No-Scale boundary conditions
means that this also provides an important mass ceiling on that parameter.

The essential broadness of the WMAP7 compliant region, all of which survives by the mechanism of stau-neutralino coannihilation,
is particularly remarkable.  One is accustomed in mSUGRA styled plots of the $(M_{1/2},M_0)$ plane to seeing
extraordinarily narrow coannihilation bands.  However, our SUSY spectrum is generated
by proportionality only to $M_{1/2}$, and thus features an extreme uniformity across the viable parameter space.  The essential
mass hierarchy $m_{\widetilde{\rm t}_{1}} < m_{\widetilde{g}} < m_{\widetilde{q}}$ of a light stop and gluino with both sparticles lighter
than all other squarks, which is responsible in particular for a quite distinctive collider
level signal of ultra-high multiplicity jet events~\cite{Li:2011hr,Maxin:2011hy},
is robust up to an overall rescaling by $M_{1/2}$.  This speaks also to the surprising ability to thread the 
WMAP7 needle so successfully and generically; more importantly however, it indicates how finely naturally adapted (not finely tuned)
No-Scale $\cal{F}$-$SU(5)$ is with regards to the question of the CDM relic density.  Its predilection for relative stability in this crucial
indicator could be a curse much more easily than a blessing, all else being equal.  Of course, the Higgs VEV ratio is also critical
to this discussion, albeit at a lower order.  With increasing $M_{1/2}$, and to a much lesser degree $M_{\rm V}$, the correspondingly
heavier bino mass must be countered by an upward shift in $\tan \beta$, which in turn decreases the neutralino annihilation cross section.  Since the
vector-like particle's contributions arise only out of the gauge coupling and gaugino mass RGEs, the dependency is weak, leading
to more rapid variations in the mass parameter $M_{\rm V}$.  This is the same mechanism which
has already been invoked to drive the transition toward a stau LSP, ultimately capping both $\tan \beta$ and $M_{\rm V}$.

%%%%%%%%%%%%%%%%%%%%%%%%%%%%%%%%%%%%%%%%%%%%%%%%%%%%%%%%%%%%%%%%

\section{The Golden Strip, Revamped\label{sct:goldenstrip}}

\begin{figure*}[htf]
        \centering
        \includegraphics[width=0.95\textwidth]{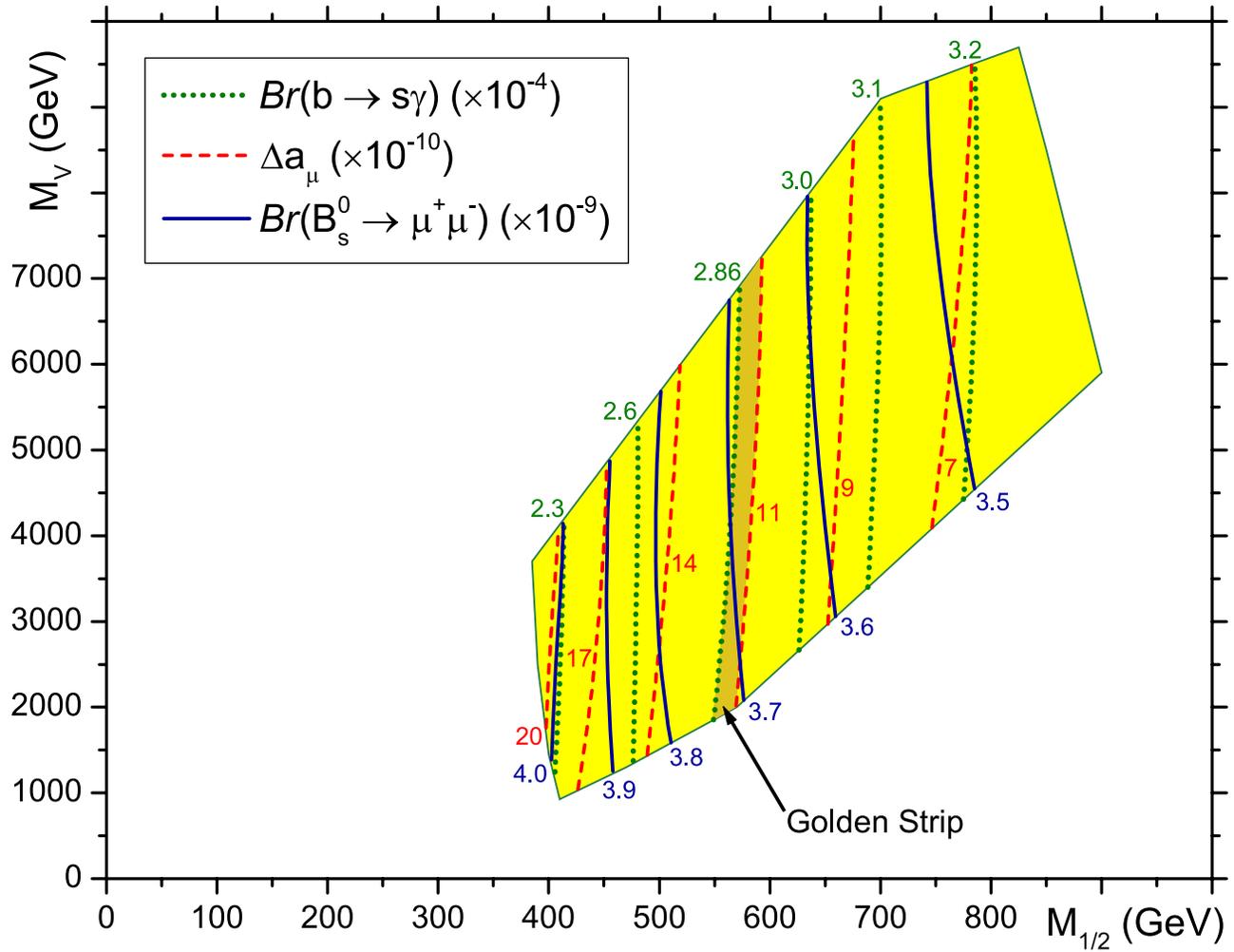}
        \caption{Contours depicting the SUSY contribution to the key rare processes responsible for shifts
in the muon anomalous moment $\Delta \left[ a_\mu \equiv (g_\mu - 2)/2 \right]$, and the FCNC decays $b \to s\gamma$
and $B_{s}^{0} \to \mu^+\mu^-$ are overlaid onto the surviving bare-minimal parameter space of No-Scale $\cal{F}$-$SU(5)$.
The region deemed to be best consistent with these overlapping constraints, featuring ${\rm Br}( b \to s \gamma) \geq 2.86 \times 10^{-4}$
and $\Delta a_\mu \geq 11 \times 10^{-10}$, is designated as the golden strip.
}
        \label{fig:Wedge_RareProcesses}
\end{figure*}

\begin{figure*}[htf]
        \centering
        \includegraphics[width=0.95\textwidth]{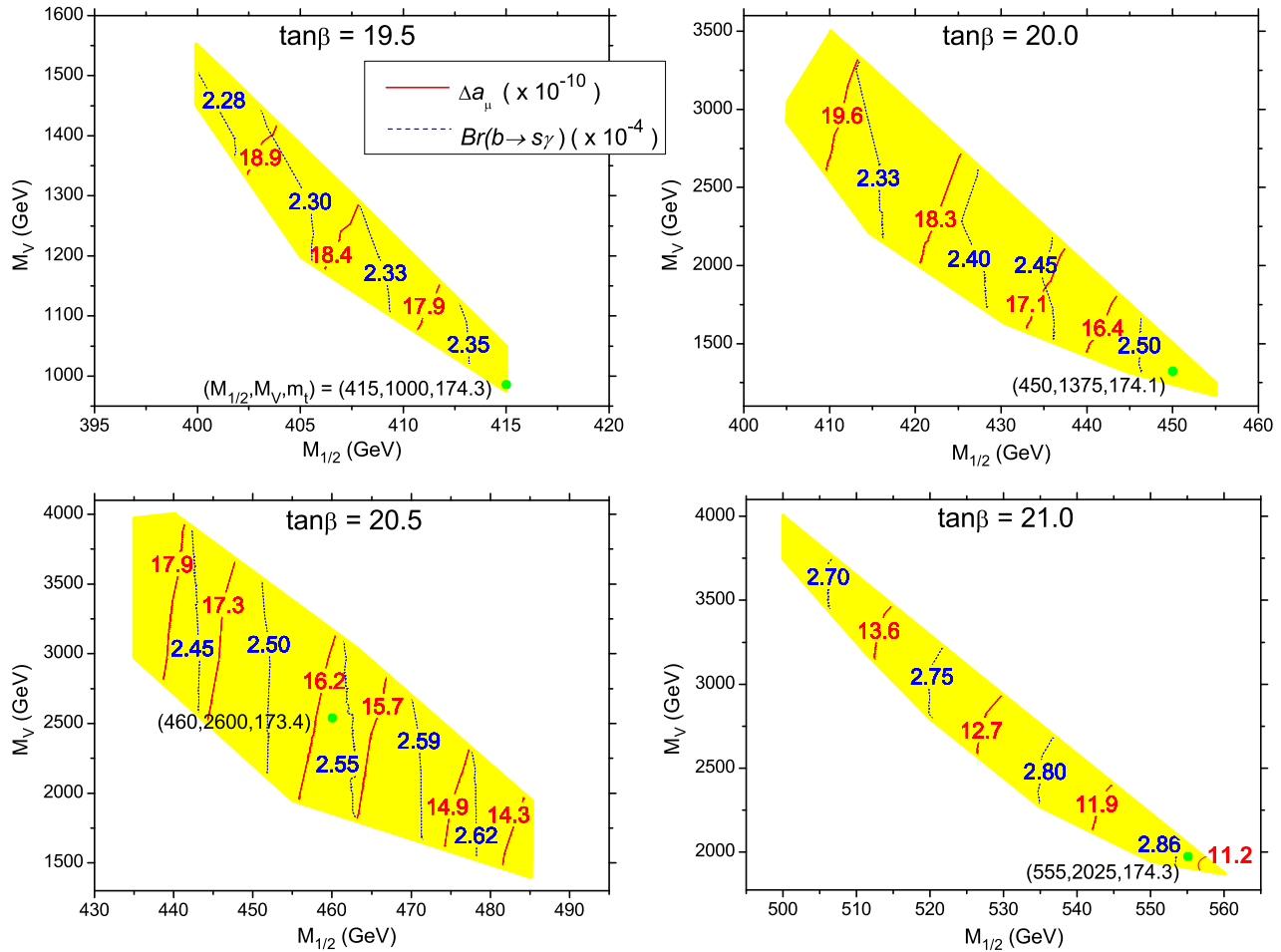}
        \caption{We display four segregated regions of the bare-minimal phenomenologically constrained parameter
space, taking the discrete values $\tan \beta = \{19.5, 20.0, 20.5, 21.0\}$.
We demarcate the contours of $\Delta a_\mu$ (red, solid) and ${\rm Br} (b \to s\gamma)$ (blue, dashed), which are not
themselves included among the bare-minimal experimental constraints, as a test of consistency with the bare-minimal constraints.
The green dots position four benchmark points selected for more detailed study,
labeled by their respective ($M_{1/2}$, $M_{\rm V}$, $m_{\rm t}$) model parameters.}
        \label{fig:exp_4plex}
\end{figure*}

There are additional phenomenological inputs, considered to be somewhat more ductile,
which have been expressly excluded from our classification of the bare-minimal constraints.
Broadly, this second category of constraints consists of limits associated with the SUSY contributions
to key rare processes, and, in particular, the flavor-changing neutral current (FCNC) decays $b \to s\gamma$
and $B_{s}^{0} \to \mu^+\mu^-$, and loops affecting the muon anomalous magnetic moment $(g-2)_\mu$.
The model subspace that is compatible with these additional criteria shall be referred to
as the ``golden strip'' of No-Scale $\cal{F}$-$SU(5)$.  The central point in our distinction is that the procedure by which
these limits are established is one of considerable intricacy, involving experimentally the subtractive measurement of higher order
corrections, supplemented for their interpretation by extremely delicate theoretical calculations.  While the stated
bounds of confidence that accompany each quoted result are certainly plausibly established, it nevertheless seems to us
that the likelihood of a future shift of the central value by more than one standard deviation, attributable in particular
to some presently unaccounted systematic effect, is substantially higher in these cases than for those already discussed.
This is not to say that we ignore these latter constraints, but rather only that their consequence will be considered and
presented in a somewhat different way.  In particular, we would like to demonstrate as a mark of phenomenological consistency
that imposing only the bare-minimal constraints produces a parameter space in which those constraints classified
as subordinate are either automatically satisfied or at least provided a non-zero intersection.
We have carefully delineated contours of the net effect for these processes in Figure~\ref{fig:Wedge_RareProcesses}.
Figure Set~\ref{fig:exp_4plex} breaks down the rare process statistics for isolated benchmark values of $\tan \beta = \{19.5, 20.0, 20.5, 21.0\}$.

Of the experiments so discussed, we have greater confidence in the pertinence of
the limits imposed by the process $b \to s\gamma$.  To be precise, any numerical discussion
will actually refer to an inclusive branching ratio for the experimentally
accessible meson transitions $\overline{B} \to X_s \gamma$ with an anti-quark spectator,
although we will employ a shorthand notation.
The results from the Heavy Flavor Averaging Group (HFAG)~\cite{Barberio:2007cr}, 
including contributions from BABAR, Belle, and CLEO, are 
${\rm Br} (b \to s\gamma) = (3.55 \pm 0.24_{\rm exp} \pm 0.09_{\rm model}) \times 10^{-4}$.
An alternate approach to the average~\cite{Artuso:2009jw}
yields a slightly lower central value, but also a smaller error, suggesting
${\rm Br} (b \to s\gamma) = (3.50 \pm 0.14_{\rm exp} \pm 0.10_{\rm model}) \times 10^{-4}$.
See Ref.\cite{Misiak:2010dz} for recent discussion and analysis.
The theoretical SM contribution at the next-to-next-to-leading order (NNLO) has been estimated
variously at ${\rm Br} (b \to s\gamma) = (3.15 \pm 0.23) \times 10^{-4}$~\cite{Misiak:2006zs}
and ${\rm Br} (b \to s\gamma) = (2.98 \pm 0.26) \times 10^{-4}$~\cite{Becher:2006pu}.
For our analysis, we will combine the quoted HFAG experimental error with the smaller of the
theoretical errors in quadrature, since there is an implicit difference taken during attribution
of the post-SM effect.  Doubling this result to establish the two standard deviation boundary yields
a net permissible error of $\pm 0.69 \times 10^{-4}$, which combines with the central
experimental value to provide limits on the simulated search range of 
$2.86 \times 10^{-4} \leq {\rm Br} (b \to s\gamma) \leq 4.24 \times 10^{-4}$.
The full model space is compliant with the upper limit, but there is some pressure
exerted by the lower limit.  Since the leading squark and gaugino
contributions to ${\rm Br} (b \to s\gamma)$ oppose the SM and Higgs terms in sign, this translates also
to a lower bound on the mass parameter $M_{1/2}$, as is seen graphically in Figure~\ref{fig:Wedge_RareProcesses}.
This occurs such that the SUSY spectrum will be sufficiently massive for suppression of the subtractive counter
terms, leaving a viable residual portion of the original SM effect.
However, the persistent difficulty in condensing this SM background out of any potential carrier of new physics
applies also to our own simulation; the version of {\tt MicrOMEGAs}~\cite{Belanger:2008sj} in current use reports
a next-to-leading order (NLO) branching ratio in the SM limit of $3.72 \times 10^{-4}$, although the most
recent production release reports adoption of a new NNLO aware algorithm which yields a SM contribution
of $3.27 \times 10^{-4}$~\cite{Belanger:2010gh}.  In conjunction with the potentially large uncertainties attributable to
the perturbative and non-perturbative QCD corrections~\cite{Misiak:2010dz}, these observations reinforce our decision to distinguish the rare process limits from the more stable bare-minimal constraints. All considered, we suggest that a substantial relaxation of the lower bound, even to the vicinity of $2.50 \times 10^{-4}$, could be plausible, reopening a majority of the bare-minimal space.

The second rare process to which we turn attention is the set of post-SM contributions to the anomalous magnetic moment of the muon,
as characterized by the difference $\Delta \left[ a_\mu \equiv (g_\mu - 2)/2 \right]$ between experiment and the calculable SM component.
The seminal measurement of $a_\mu$ was completed several years back by experiment E821 at Brookhaven National Laboratory, employing
the Alternating Gradient Synchrotron~\cite{Bennett:2004pv}.  The reported experimental value is $a_\mu^{\rm exp} = (11,659,208 \pm 6) \times 10^{-10}$,
precise to approximately half a part per million.  Curiously, the seeding of a theoretical calculation with $e^+e^-$ annihilation data
is reported~\cite{Bennett:2004pv} to yield a result $a_\mu^{{\rm th} (e^+e^-)} = (11,659,181 \pm 8) \times 10^{-10}$ that is only
marginally consistent with the corresponding result $a_\mu^{{\rm th} (\tau)} = (11,659,181 \pm 8) \times 10^{-10}$ seeded by data from $\tau$ decays.
Combining errors in quadrature, and doubling to the two standard deviation level, the resulting differences are
$\Delta a_\mu^{e^+e^-} = 27 \pm 20 \times 10^{-10}$ and $\Delta a_\mu^{\tau} = 12 \pm 18 \times 10^{-10}$, respectively.
More recent analyses instead yield $\Delta a_\mu = 25.9 \pm 16.2 \times 10^{-10}$~\cite{Cho:2011rk} and,
$\Delta a_\mu = 26.1 \pm 16.0 \times 10^{-10}$~\cite{Hagiwara:2011af}.  As before, the model is in no danger from the upper limit,
although the lower limit may again be in play.  However, in this case the post SM contributions to $\Delta a_\mu $ are instead additive,
so that a lower bound on the net effect translates instead to an upper bound on $M_{1/2}$.  Working in opposition, these constraints might be
interpreted to create a narrow region of preferred phenomenology.  Conservatively, we might elect to enforce $\Delta a_\mu \geq 11$, corresponding to
generation of the golden strip highlighted in Figure~\ref{fig:Wedge_RareProcesses}.  This would correspond to a gaugino boundary mass $M_{1/2}$
within the approximate range of $(560\text{--}600)$~GeV.  However, the $\Delta a_\mu$ bound is one to which we extend somewhat less credulity,
and it is not difficult to argue for a value of 9 or 10 (or possibly even less), again substantially expanding the favored window.

The final rare processes to be considered are, like $b \to s\gamma$, decays proceeding via a flavor-changing neutral intermediary.
In particular, we are referring to $B^0_{s,d} \to \mu^+\mu^-$, where the initial quark content may be either $(s,\overline{b})$,
or $(d,\overline{b})$.  Omission of the subscript implies the latter case, which features an experimental
upper bound on the branching ratio that is stronger (smaller) by almost a magnitude order.
The SM expectation for the branching ratios of $B^0_{s,d} \to \mu^+\mu^-$ are $3.2 \pm 0.2 \times 10^{-9}$
and $1.1 \pm 0.1 \times 10^{-10}$ respectively~\cite{Buras:2010wr}, where the loop-level process employs a
virtual $W$-boson to transmute the quark content, facilitating a $t\bar{t} \to Z^0$ fusion event.
The CMS upper bounds on these processes are $1.9 \times 10^{-8}$ and $4.6 \times 10^{-9}$, based on $1.14~{\rm fb}^{-1}$ of data~\cite{Chatrchyan:2011kr}.
The corresponding LHCb upper bounds are $1.4 \times 10^{-8}$ and $3.2 \times 10^{-9}$, based on $0.41~{\rm fb}^{-1}$ of data~\cite{LHCb:2011ac}.
These experiments have both already eclipsed the best limits from the Tevatron.  Curiously, CDF is the only collaboration, based upon about
$7~{\rm fb}^{-1}$ of data, to claim an observed excess sufficient to establish a lower bound, quoted as
${\rm Br} (B^0_{s} \to \mu^+\mu^-) \geq 4.6 \times 10^{-9}$~\cite{Aaltonen:2011fi}; however, the central value of the reported
observation is slightly in excess of the LHCb upper bound.  All together, we will follow the lead of Ref.~\cite{Stone:2011vd},
which recognizes a combined CMS/LHCb limit of Br($B^0_{s} \to \mu^+\mu^-) \leq 1.1 \times 10^{-8}$.  Since the SUSY rate
for this process is proportional to a sixth power of $\tan \beta$, and since No-Scale $\cal{F}$-$SU(5)$ globally enforces
a rather low value of $\tan\beta \leq 23$, the model space is in no jeopardy from this bound.

A readily apparent consequence of our present procedural refinements is visible in the shifting location
of the Figure~\ref{fig:Wedge_RareProcesses} golden strip, driven by the ${\rm Br} (b \to s \gamma)$ limits
toward a somewhat heavier gaugino mass $M_{1/2}$ than was predicted by our initial effort~\cite{Li:2010mi}.
The linked model dependencies embodied in the steeply inclined phenomenologically allowed bare-minimal region
likewise enforces a somewhat larger $\tan\beta$ and a possibly substantially larger vector-like mass $M_{\rm V}$.
As will be further elaborated in Section~(\ref{sct:acount}), the same characteristics which make No-Scale $\cal{F}$-$SU(5)$
non-trivially predictive (in a manner approaching over-constraint) may conversely imply that certain numerical outputs
will be geared for a rather sensitive response to changes elsewhere in the model.
We emphasize that these numerical offsets should be conceptually decoupled from the presentation of the dramatically
enlarged ``bare-minimal'' parameter space in Section~(\ref{sct:baremin}), which is an entirely new construct, presented
for the first time in the current paper.  The rare-process restricted channel (the golden strip) of Figure~(\ref{fig:Wedge_RareProcesses})
may be fairly compared with the previously advertised golden strip~\cite{Li:2010mi}, and the subspace further restricted
by an externally defined value of $M_{\rm V} = 1000$~GeV may be fairly compared to the previous ``golden point''~\cite{Li:2010ws}.
When comparing in this manner predictions established under common input assumptions, the sizes of the respective parameter
spaces and the basic phenomenological character of the solutions remains essentially unchanged; the only modifications are
absolute shifts in the numerical fitting, again wholly attributable to i) an improvement in the implementation of the matching
condition on $B_\mu$, and ii) a correction of rounding errors in the $\beta$-function coefficients of the vector-like multiplet RGEs.
We thus strongly stand by the integrity of the results given in Refs.~\cite{Li:2010ws,Li:2010mi}, within the context of
the model assumptions in play at that time.

The elevation of $M_{\rm V}$ above the boundaries which we have previously seriously entertained is of some concern.
One notable issue is that the vector-like particle mass $M_{\rm V}$, in certain regions of the
bare-minimal constraint parameter space, becomes so large that we cannot hope to observe such particles even at the
future ${\sqrt s}=14$~TeV LHC run.  However, this is purely a complaint of convenience, and not a physical argument against.
A more substantive objection certainly exists against the new hierarchy problem which would emerge if $M_{\rm V}$
were elevated substantially out of the electroweak order, but this is only suggestive, rather than strictly predictive.
Although wary of the prospect of letting the baseline vector-like mass grow by anything approaching a full order of magnitude,
we see no objection of principle against a measured elevation of $M_{\rm V}$ which keeps it broadly of the electroweak order.
Viewed logarithmically, as is of course appropriate in light of the
renormalization group structure, a shift by a multiplicative factor of $2 - 4$ is not outrageous.
The rare-process constraints, particularly those on $\Delta a_\mu$ may be phenomenologically helpful in this regard.
This is moreover consistent with the original motivation of No-Scale GUTs, since
the Super No-Scale condition itself becomes quite subtle if the vector-like particle mass is much larger
than the sparticle masses~\cite{Cremmer:1983bf,Ellis:1983sf,Ellis:1983ei,Ellis:1984bm,Lahanas:1986uc}.
Incidentally, we have considered the possibility of a contribution to the rare processes by the vector-like multiplets
themselves, but quickly concluded that the extreme loop-level mass-squared suppression would render
their participation comparatively irrelevant.

%%%%%%%%%%%%%%%%%%%%%%%%%%%%%%%%%%%%%%%%%%%%%%%%%%%%%%%%%%%%%%%%

\section{The Super No-Scale Mechanism\label{sct:minmin}}

As a check of broad compatibility with the top-down theoretical perspective, we select a discrete region of the bottom-up
phenomenological bare minimal model space for further study under application of the ``Super No-Scale'' condition,
as studied in two prior works~\cite{Li:2010uu,Li:2011dw}.  However, in the spirit of the bare-minimal constraints,
this theoretical augmentation is of the most generic possible variety.  Specifically, this procedure compares the
minimum $V_{EW}^{\rm min}$ of the scalar Higgs potential (after consistent electroweak symmetry breaking is enforced)
along a continuously connected string of adjacent model parameterizations, dynamically selecting out 
the model with the smallest locally bound value of $V_{EW}^{\rm min}$.
This point of secondary minimization is referred to as the ``minimum minimorum''.
Note that any numerical values given in
plots refer in actuality to the signed fourth root of the Higgs potential, in units of GeV.

For momentarily fixed values of $m_{\rm t}$ and $M_{\rm V}$, one
recognizes that the two EWSB minimization conditions may be taken first to determine the Higgs bilinear mass term
$\mu$ at $M_{\cal F}$, and since $B_\mu(M_{\cal F}) = 0$ is already fixed by the No-Scale boundary conditions,
secondly to establish $\tan\beta$ as an implicit function of the universal gaugino boundary mass $M_{1/2}$.  The
resulting continuous string of minima of the broken Higgs potential $V_{EW}^{\rm min}$ are then likewise labeled by their
value of $M_{1/2}$.  Because the minimum of the electroweak Higgs potential 
$V_{EW}^{\rm min}$ depends on the gaugino mass $M_{1/2}$, and $M_{1/2}$ 
is in turn related to the F-term of the K\"ahler modulus $T$ in the weakly coupled heterotic
$E_8 \times E_8$ string theory or M-theory on $S^1 / Z_2$, 
the gaugino mass is determined by the equation $d V_{EW}^{\rm min} /dM_{1/2}=0$ in correspondence with
the modulus stabilization~\cite{Ellis:1983sf,Lahanas:1986uc}.
At this locally smallest value of $V_{EW}^{\rm min} (M_{1/2})$, {\it i.e.}~ 
the minimum minimorum, the dynamic determination of $M_{1/2}$, as well as the
parametrically coupled value of $\tan \beta$, is established.

\begin{figure*}[htf]
        \centering
        \includegraphics[width=0.97\textwidth]{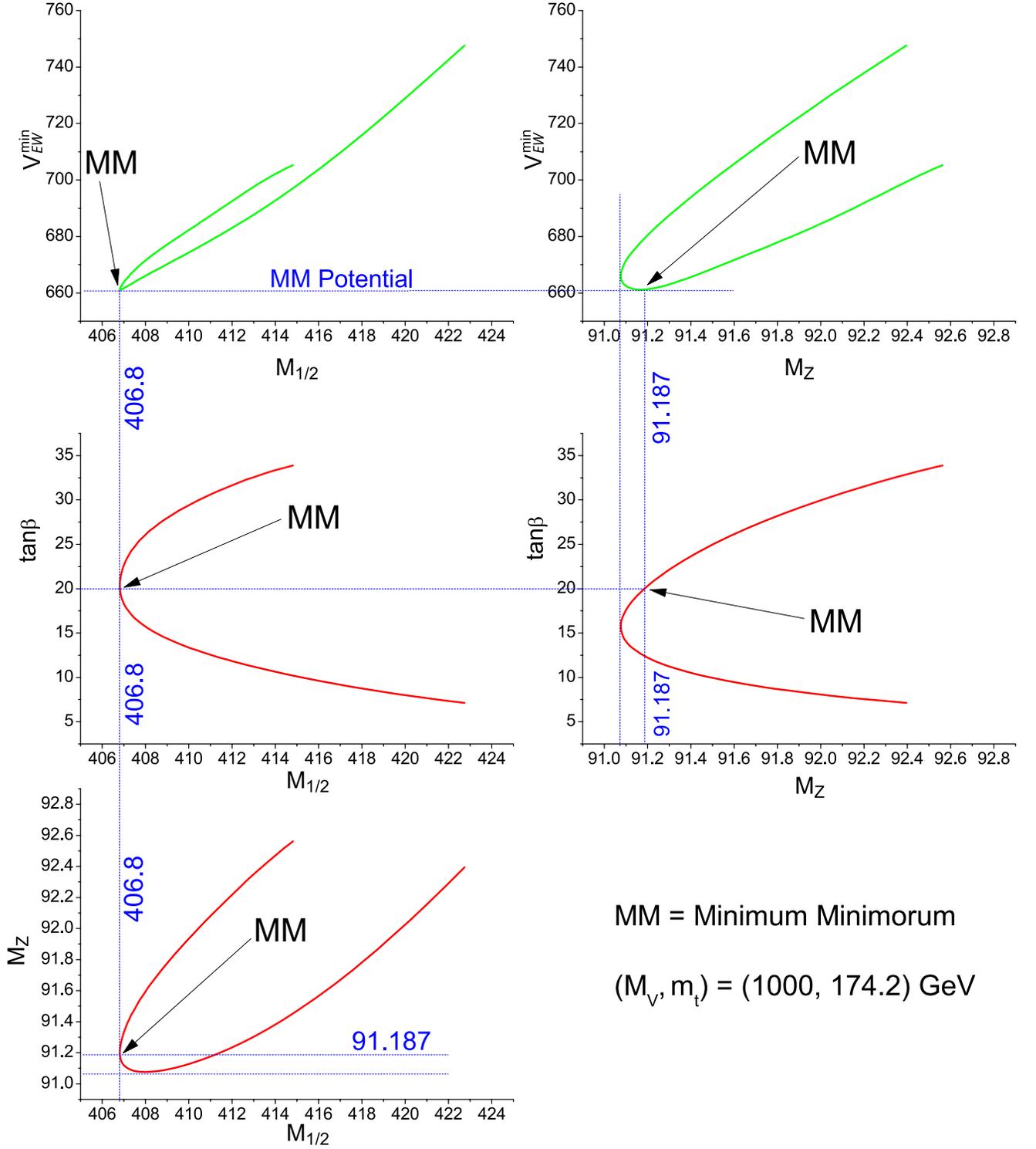}
        \caption{We depict the correlation of model parameters facilitating a dynamical determination of the EW scale.
Annotated on each curve is the minimum minimorum, defined as the secondary minimization of a continuously connected
string of EWSB minima $V_{EW}^{\rm min}$ of the 1-loop Higgs potential, at which the physical vacuum will be localized.
The example illustrates $M_{\rm Z}$ = 91.187 as the dynamically determined electroweak scale in
No-Scale $\cal{F}$-$SU(5)$.  The minimum minimorum occurs near the minimum value of the modulus $M_{1/2}$, primarily as a
consequence of the heavy squark 1-loop contributions to the Higgs potential.  The vector and top quark mass here are
fixed at ($M_{\rm V}, m_{\rm t}$) = (1000,174.2)~GeV for computation of the Higgs potential.}
        \label{fig:Vmin_5plex}
\end{figure*}

\begin{figure*}[htf]
        \centering
        \includegraphics[width=0.95\textwidth]{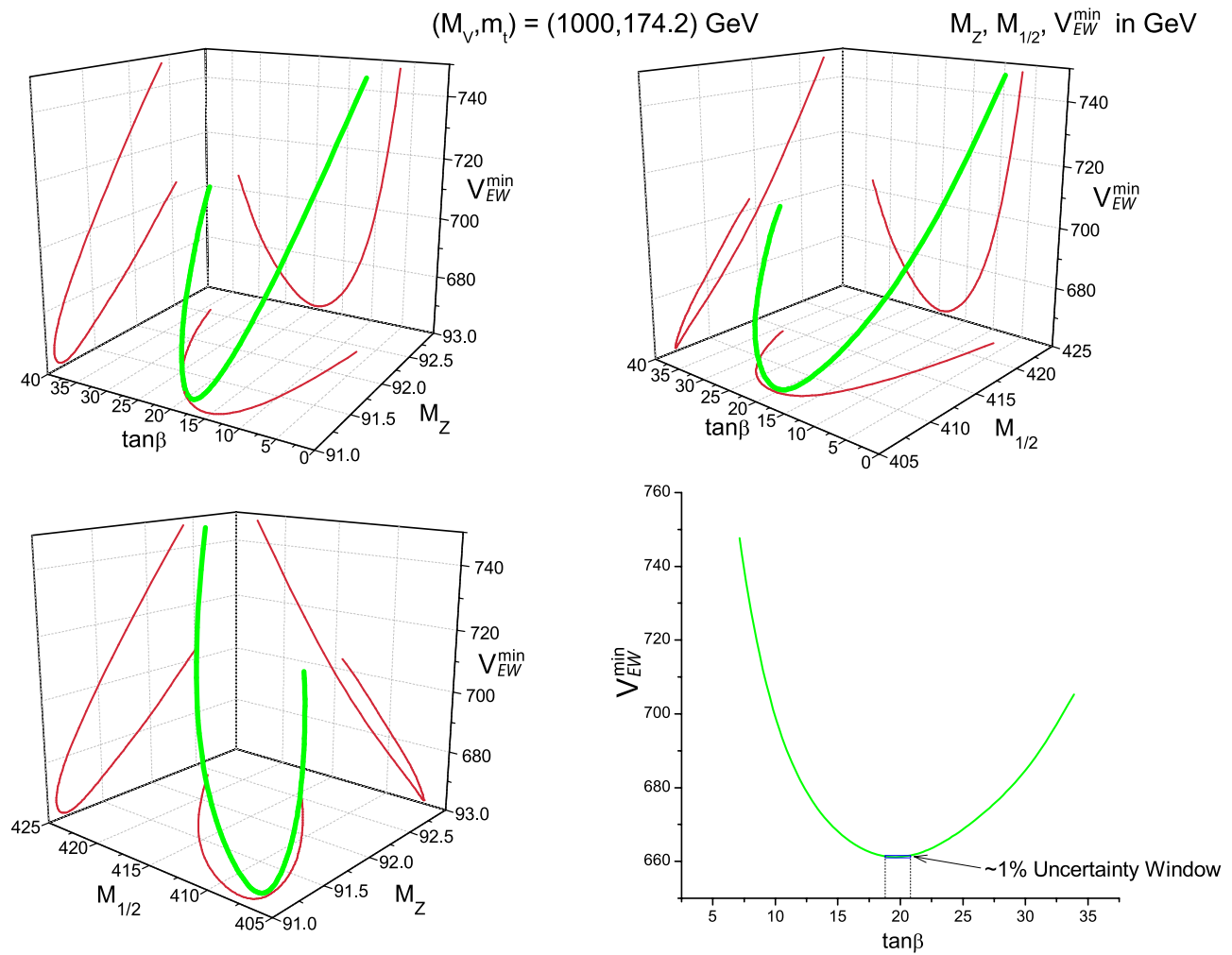}
        \caption{Three-dimensional (3D) illustrations of the secondarily minimized 1-loop Higgs potential exhibited in
Figure~\ref{fig:Vmin_5plex}. The green curves existing in each respective 3D space are the 1-loop Higgs potential, while
the red curves embedded within the three flat mutually perpendicular planes are the projections of the Higgs potential onto
these smooth planes. The lower right curve is an extraction of ($V_{EW}^{\rm min}$, $\tan \beta$) from the ($M_{\rm Z}$, $\tan \beta$, $V_{EW}^{\rm min}$)
and ($M_{1/2}$, $\tan \beta$, $V_{EW}^{\rm min}$) 3D spaces. The ($V_{EW}^{\rm min}$, $\tan \beta$) window of uncertainty is outlined, distinguished by
an approximate 1\% deviation of the Higgs potential $V_{EW}^{\rm min}$ from the precise numerical minimum minimorum, comparable in scale to the
QCD corrections to the Higgs potential at the second loop.}
        \label{fig:Vmin_4plex}
\end{figure*}

For the present study we favor the extension of this technique advanced in our more recent
treatment~\cite{Li:2011dw}, and demonstrated graphically in Figure Sets~\ref{fig:Vmin_5plex}--\ref{fig:Vmin_4plex}.  The
key distinction is that we allow fluctuation not only of $M_{1/2}$, but also of the GUT scale Higgs modulus as embodied in the
mass scale $M_{32}$ at which the $SU(3) \times SU(2)_{\rm L}$ couplings initially meet.  In actual practice, the variation of
$M_{32}$ is achieved in the reverse by programmatic variation of the Weinberg angle, holding the strong and electromagnetic couplings at
their physically measured values; this is achieved in turn by fluctuation of the $Z$-boson mass, the magnitude
of the Higgs VEV being held essentially constant.  Simultaneous to the recognition of the presence of a second
dynamic modulus, we must lock down the value of $\mu$, which by contrast is a simple numerical parameter, and
ought then to be treated in a manner consistent with the top quark and vector-like mass parameters.
Since two dynamic constraints ($B_\mu = 0$ and $\mu = {\rm constant}$) are enforced during the comparison, the string of
model parameterizations must transit a three-dimensional scanning volume.  We thus consider that, for fixed $M_{\rm V}$
and $m_{\rm t}$, the value of $V_{EW}^{\rm min}$ may be compared among interconnected (singly parameterized)
triplets of the free parameters $M_{1/2}, \tan\beta,~{\rm and}~M_{\rm Z}$, dynamically selecting a single such combination
of all three parameters.  For our example, we choose $M_{\rm V} = 1000$~GeV and $m_{\rm t} = 174.2$~GeV, easily within the region
where the vector-like particles should be able to be produced at the future LHC.

Interestingly, by extracting in this manner a constant $\mu$ slice of the $V_{EW}^{\rm min}$ hyper-surface,
the secondary minimization condition on $\tan \beta$ is effectively shifted to a somewhat larger value.
In our example, the dynamically selected value of $\tan \beta$ is very close to $20$, and in excellent
parametric agreement with a point near the lower left edge of the bare-minimal model space.  This striking 
demonstration of compatibility with the bottom-up approach is a key result.  In particular, the rather stable value of
$\tan \beta$ in the vicinity of 20 which is phenomenologically imposed on the model as a whole presents a rather narrow target for
the dynamic determination, and much more so when matching specific fixed values of $M_{\rm V}$ and $m_{\rm t}$.

Curiously, we find (for fixed Z-Boson mass) that the gaugino mass $M_{1/2}$ is almost equal to the Higgs bilinear
mass term $\mu$ across the entire phenomenologically allowed region. This may be an effect of the
strong No-Scale boundary conditions and might further have implications to the solution of the $\mu$ problem
in the supersymmetric standard model~\cite{LMNW-P}.  The fact that $\mu$ and $M_{\rm V}$ might be generated from
the same mechanism~\cite{LMNW-P} represents an additional naturalness argument for the suggestion that
$\mu$ and $M_{\rm V}$ should be of the same order.  A broader verification of the compatibility of the phenomenological
(driven by the CDM relic density) and theoretical (driven by the Super No-Scale condition) perspectives, as embodied in the
consistency of $\tan \beta$ and $M_{1/2}$ for various $M_{\rm V}$, $m_{\rm t}$, and $\mu$ groupings,
constitutes a separate study~\cite{Li:2011ex}.

%%%%%%%%%%%%%%%%%%%%%%%%%%%%%%%%%%%%%%%%%%%%%%%%%%%%%%%%%%%%%%%%

\section{Additional Phenomenology}

There is another key experimental result to which we have devoted considerable attention in the past, and for which the contributions of the
vector-like fields, and specifically the dramatic increase in the $SU(3)_C \times SU(2)_L$ unified 
 coupling $g_{32}$, are particularly germane.
We refer to the current $90\%$ confidence level lower bounds on the proton lifetime of $8.2\times 10^{33}$ and $6.6\times 10^{33}$ Years
for the leading $p\to e^+ \pi^0$ and $p\to \mu^+ \pi^0$ modes~\cite{:2009gd}.  These results, which preliminary data updates now suggest
may actually be pushed into the low $10^{34}$~year order, have been compiled by the 50-kiloton (kt) water \v{C}erenkov detector at the
Super-Kamiokande facility.  They are rather clearcut, having no competing background for detection.
However, we find that they do not provide any appreciable reduction of the current parameter space.  Indeed, the predicted lifetime
for the majority of the WMAP7 region sits coyly just outside the experimental limit.

Although featuring some dependence on $\tan \beta$, the central partial lifetime for proton decay in the ${(e|\mu)}^{+} \pi^0 $ channels
falls around $(3\text{--}4) \times 10^{34}$ years, certainly testable at the future Hyper-Kamiokande~\cite{Nakamura:2003hk} and DUSEL~\cite{Raby:2008pd}
experiments.  This presently safe, yet characteristically fast and imminently observable proton lifetime, is a rather
stable feature of our model which we have studied extensively~\cite{Li:2009fq,Li:2010dp,Li:2010rz}.
Incidentally, the very slight downward shift which may be recognizable in our central proton lifetime may be attributed chiefly to the fact 
that we have now opted to substitute our former proprietary treatment of the relevant RGEs for that
provided directly by {\tt SuSpect 2.34}~\cite{Djouadi:2002ze}.  The comparatively meager fluctuation in the proton lifetime
produced by this condensation of our calculational strategy, especially in light of the extremely strong dependencies embedded
in this statistic (quartic proportionality to the $SU(3)_C \times SU(2)_L$ unified scale $M_{32}$, 
and inverse-quartic for the unified coupling $g_{32}$ measured at that scale) reinforces our confidence in the basic consistency of the numerics.

\begin{figure*}[htf]
        \centering
        \includegraphics[width=0.95\textwidth]{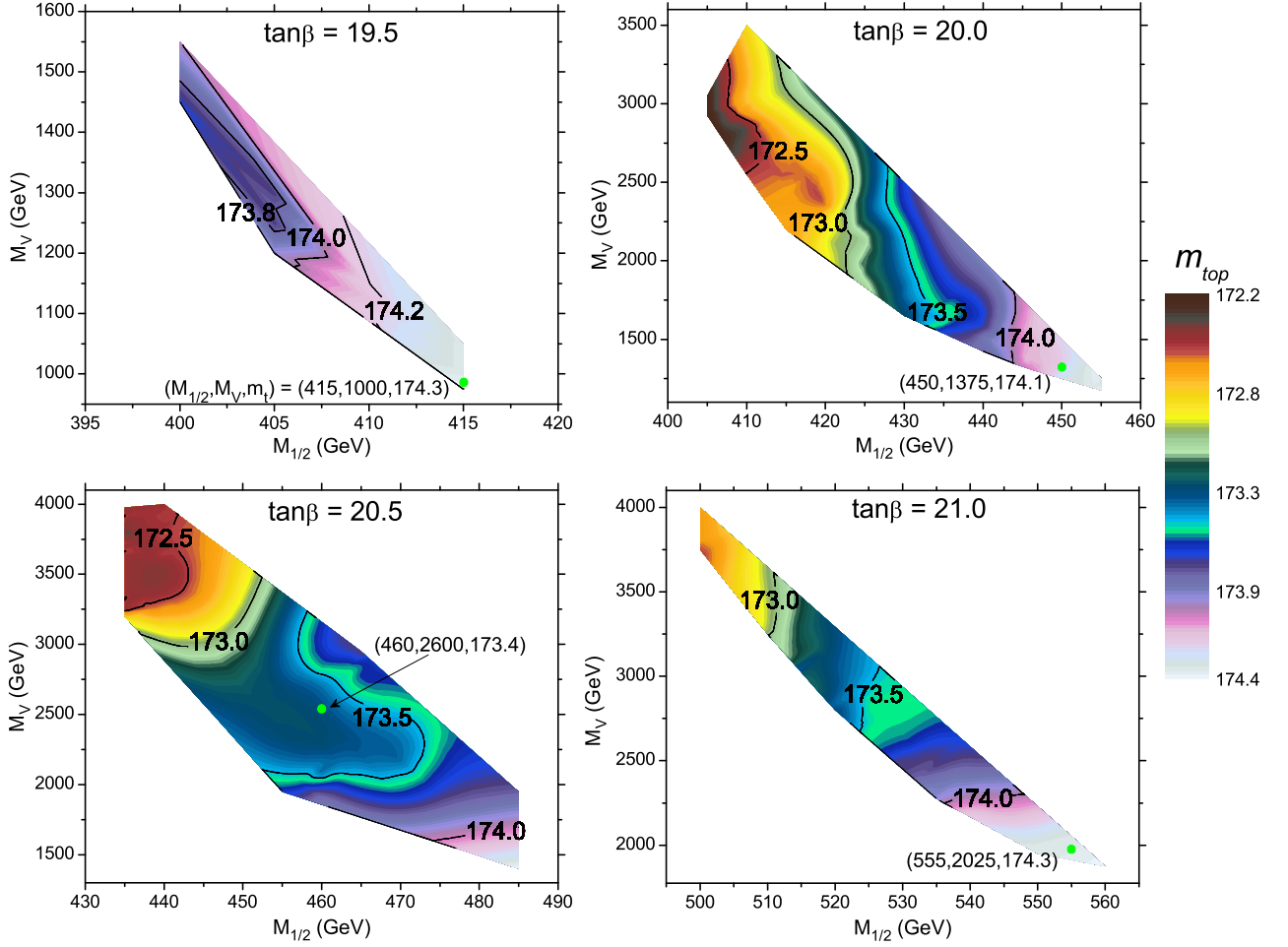}
        \caption{Contours of the top quark mass $m_{\rm t}$ for the four segregated regions of the bare-minimal phenomenologically constrained parameter space.
The top mass is a member of the base set of bare-minimal experimental constraints; we maintain strict adherence to the world average 172.2 $\leq m_{\rm t} \leq$ 174.4.
The green dots position the four chosen benchmark points,
labeled by their respective ($M_{1/2}$, $M_{\rm V}$, $m_{\rm t}$) model parameters.
The legend associates the shading color with a numerical value of the top quark mass.}
        \label{fig:mtop_4plex}
\end{figure*}

\begin{figure*}[htf]
        \centering
        \includegraphics[width=0.95\textwidth]{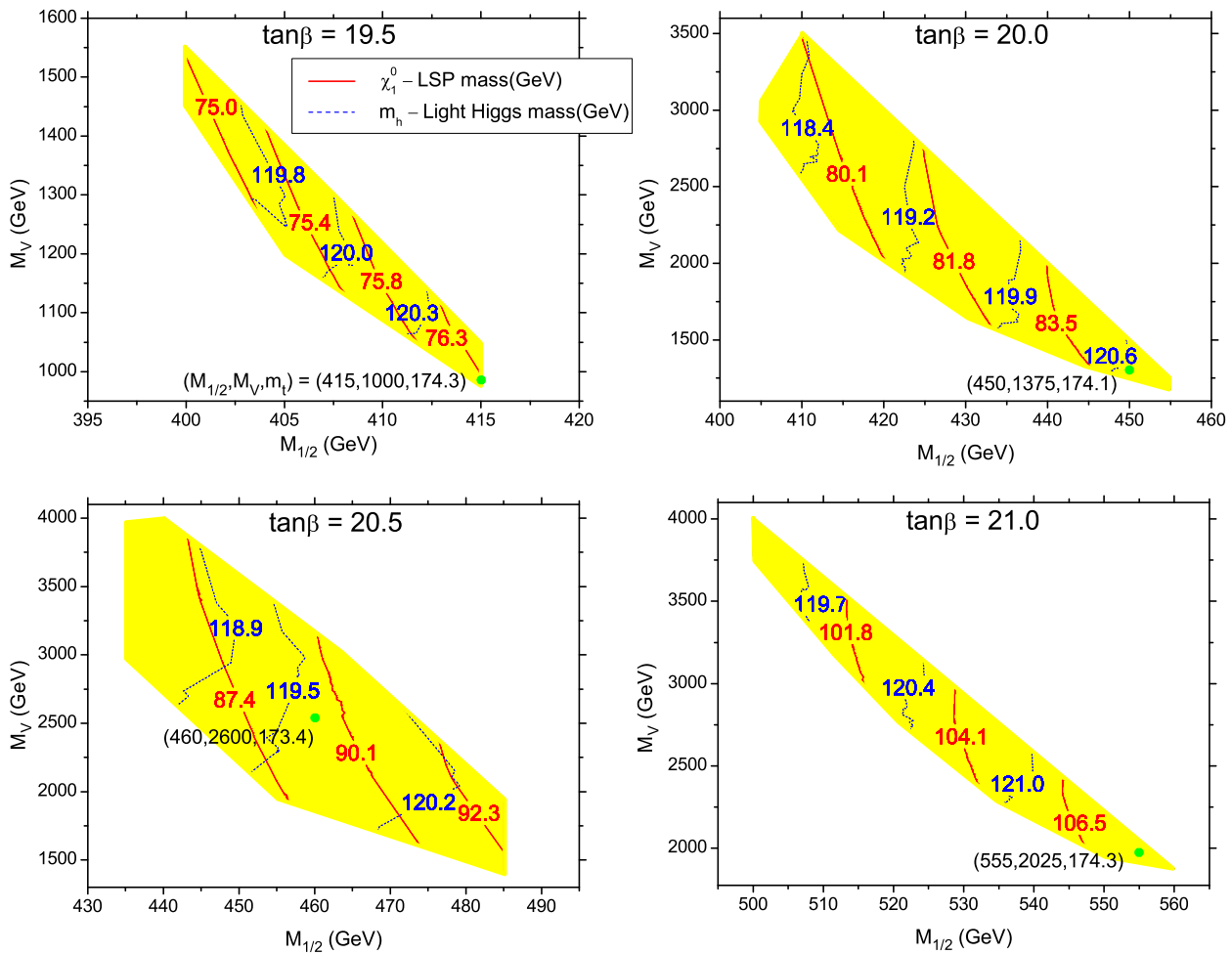}
        \caption{Contours of the lightest supersymmetric particle $\widetilde{\chi}_{1}^{0}$ mass (red, solid) and CP-Even Higgs boson mass $m_{h}$ (blue, dashed)
in GeV for the four segregated regions of the bare-minimal phenomenologically constrained parameter space.
We emphasize that $m_{h} \simeq 120$~GeV for the phenomenologically constrained parameter space~\cite{Li:2011xg}, not including an upward shift
of 3--4~GeV which may be induced by radiative loops in the vector-like fields~\cite{Li:2011ab}.
The green dots position the four chosen benchmark points, labeled by their respective ($M_{1/2}$, $M_{\rm V}$, $m_{\rm t}$) model parameters.}
        \label{fig:LSP_4plex}
\end{figure*}

\begin{figure*}[htf]
        \centering
        \includegraphics[width=0.95\textwidth]{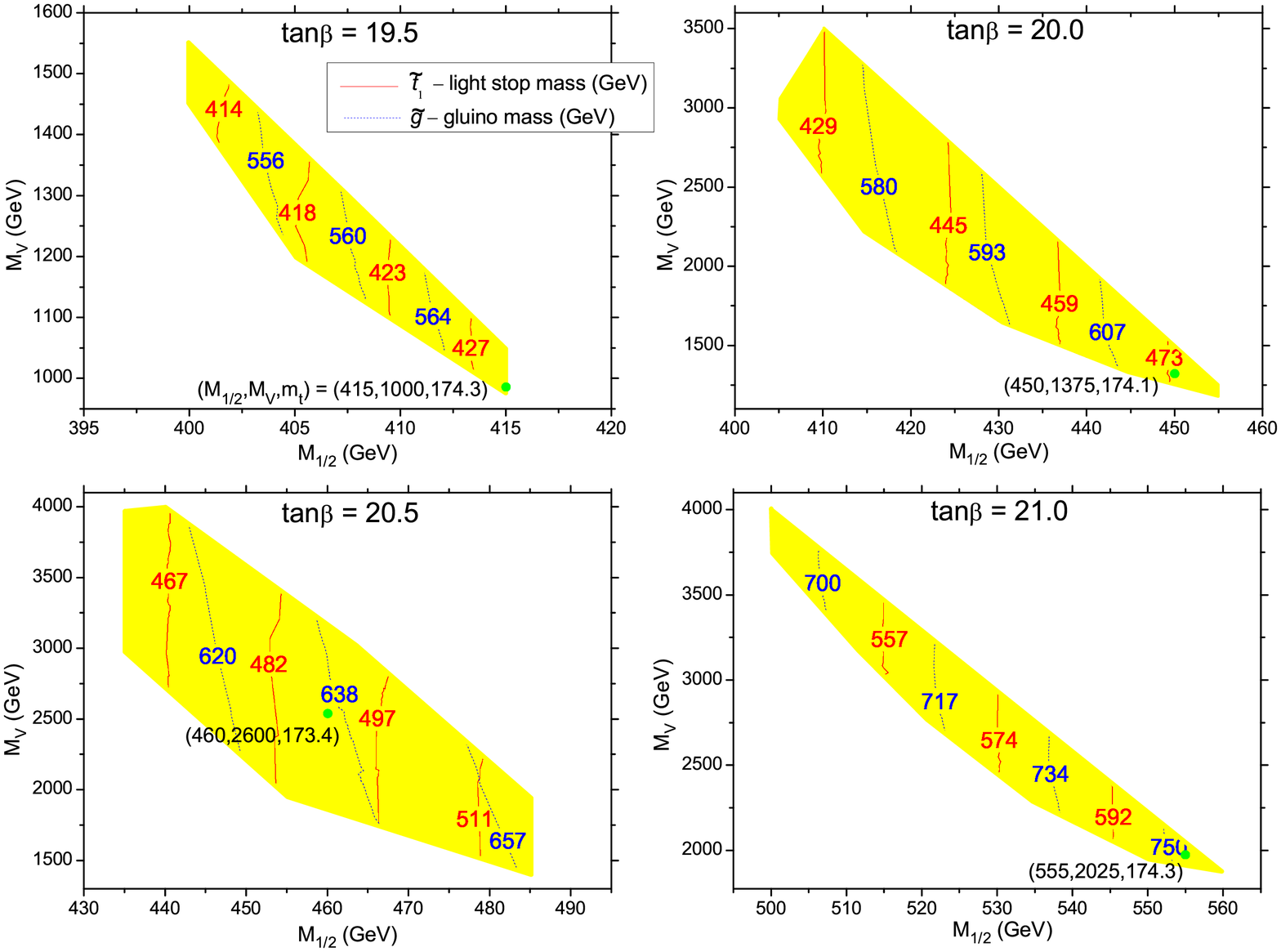}
        \caption{Contours of the light stop $\widetilde{\rm t}_{1}$ mass (red, solid) and gluino $\widetilde{g}$ mass (blue, dashed) in GeV for the four segregated regions of the
bare-minimal phenomenologically constrained parameter space.  We choose to graphically accentuate the mass differential between the
$\widetilde{\rm t}_{1}$ and $\widetilde{g}$ due to the unique and significant nature of the relationship between the
gluino and squarks in No-Scale $\cal{F}$-$SU(5)$.  Experimental validation of the $m_{\widetilde{\rm t}_{1}} < m_{\widetilde{g}} < m_{\widetilde{q}}$ mass
hierarchy would provide striking evidence for the existence of a No-Scale $\cal{F}$-$SU(5)$ vacuum. The green dots position the four chosen benchmark points,
labeled by their respective ($M_{1/2}$, $M_{\rm V}$, $m_{\rm t}$) model parameters.}
        \label{fig:stop_4plex}
\end{figure*}

We choose a sampling of representative benchmark points which span the region of bare-minimal constraints, adhering also with
some varied level of devotion to the subordinate phenomenological conditions.  These points are
dotted in green within Figure Sets~\ref{fig:exp_4plex}--\ref{fig:stop_4plex}, representing the discrete values $\tan \beta = \{19.5, 20.0, 20.5, 21.0\}$.
The detailed sparticle and Higgs spectra of each benchmark are presented in Tables~\ref{tab:masses_1}-\ref{tab:masses_4}. 
The LSP is almost entirely bino across the model space, featuring a spin-independent annihilation cross section $\sigma_{SI}$ around $2\times 10^{-10}$~pb,
presenting an excellent candidate for near term direct detection by the Xenon collaboration~\cite{Aprile:2011hi}, which
has some realistic hopes of trumping the collider based search for signs of supersymmetry at the LHC and the Tevatron~\cite{Li:2011in}.
We additionally provide the photon-photon annihilation cross-section $\left\langle \sigma v \right\rangle_{\gamma\gamma}$ with each table,
for comparison with the Fermi-LAT Space Telescope~\cite{Abdo:2010dk} results.

In Figure Set~\ref{fig:mtop_4plex}, we exhibit detailed contours of the top quark mass within the surviving parameter space for the four selected values of $\tan \beta$.
As described, we limit our freedom for the top mass input to the experimental world average of $m_{\rm t}$ = 173.3 $\pm$1.1 GeV. Likewise, Figure Set~\ref{fig:LSP_4plex}
highlights the mass contours of the LSP and light Higgs. We wish to accentuate the stability of the light Higgs mass locally around 120~GeV~\cite{Li:2011xg}.
This prediction is exclusive of radiative loop corrections from the vector-like multiplets, which may create an upward shift of 3--4~GeV~\cite{Li:2011ab}.
In addition, we show the relationship between the light stop $\widetilde{\rm t}_{1}$ and the gluino $\widetilde{g}$ in Figure Set~\ref{fig:stop_4plex}.
A remarkable consequence of this $m_{\widetilde{\rm t}_{1}} < m_{\widetilde{g}} < m_{\widetilde{q}}$ mass hierarchy is strong production
of ultra-high multiplicity ($\geq 9$) jet events at the LHC.

\begin{table}[ht]
  \small \centering
	\caption{Spectrum (in GeV) for the $\tan \beta$ = 19.5 benchmark point. 
Here, $M_{1/2}$ = 415 GeV, $M_{\rm V}$ = 1000 GeV, $m_{\rm t}$ = 174.3 GeV, $M_{\rm Z}$ = 91.187 GeV, $\Omega_{\chi}$ = 0.1139, $\sigma_{SI} = 3.1 \times 10^{-10}$ pb,
and $\left\langle \sigma v \right\rangle_{\gamma\gamma} = 5.6 \times 10^{-28} ~cm^{3}/s$. The central prediction for the $p \!\rightarrow\! {(e\vert\mu)}^{\!+}\! \pi^0$
proton lifetime is around $3.2 \times 10^{34}$ years. The lightest neutralino is 99.8\% bino.}
		\begin{tabular}{|c|c||c|c||c|c||c|c||c|c||c|c|} \hline		
    $\widetilde{\chi}_{1}^{0}$&$76$&$\widetilde{\chi}_{1}^{\pm}$&$167$&$\widetilde{e}_{R}$&$159$&$\widetilde{\rm t}_{1}$&$429$&$\widetilde{u}_{R}$&$875$&$m_{h}$&$120.4$\\ \hline
    $\widetilde{\chi}_{2}^{0}$&$167$&$\widetilde{\chi}_{2}^{\pm}$&$764$&$\widetilde{e}_{L}$&$474$&$\widetilde{\rm t}_{2}$&$830$&$\widetilde{u}_{L}$&$949$&$m_{A,H}$&$823$\\ \hline
     $\widetilde{\chi}_{3}^{0}$&$759$&$\widetilde{\nu}_{e/\mu}$&$467$&$\widetilde{\tau}_{1}$&$86$&$\widetilde{b}_{1}$&$771$&$\widetilde{d}_{R}$&$910$&$m_{H^{\pm}}$&$829$\\ \hline
    $\widetilde{\chi}_{4}^{0}$&$763$&$\widetilde{\nu}_{\tau}$&$457$&$\widetilde{\tau}_{2}$&$467$&$\widetilde{b}_{2}$&$874$&$\widetilde{d}_{L}$&$953$&$\widetilde{g}$&$567$\\ \hline
		\end{tabular}
		\label{tab:masses_1}
\end{table}

\begin{table}[ht]
  \small \centering
	\caption{Spectrum (in GeV) for the $\tan \beta$ = 20.0 benchmark point. 
Here, $M_{1/2}$ = 450 GeV, $M_{\rm V}$ = 1375 GeV, $m_{\rm t}$ = 174.1 GeV, $M_{\rm Z}$ = 91.187 GeV, $\Omega_{\chi}$ = 0.1155, $\sigma_{SI} = 2.5 \times 10^{-10}$ pb,
and $\left\langle \sigma v \right\rangle_{\gamma\gamma} = 4.4 \times 10^{-28} ~cm^{3}/s$. The central prediction for the $p \!\rightarrow\! {(e\vert\mu)}^{\!+}\! \pi^0$
proton lifetime is around $3.6 \times 10^{34}$ years. The lightest neutralino is 99.8\% bino.}
		\begin{tabular}{|c|c||c|c||c|c||c|c||c|c||c|c|} \hline		
    $\widetilde{\chi}_{1}^{0}$&$85$&$\widetilde{\chi}_{1}^{\pm}$&$185$&$\widetilde{e}_{R}$&$172$&$\widetilde{\rm t}_{1}$&$474$&$\widetilde{u}_{R}$&$930$&$m_{h}$&$120.6$\\ \hline
    $\widetilde{\chi}_{2}^{0}$&$185$&$\widetilde{\chi}_{2}^{\pm}$&$804$&$\widetilde{e}_{L}$&$504$&$\widetilde{\rm t}_{2}$&$878$&$\widetilde{u}_{L}$&$1010$&$m_{A,H}$&$867$\\ \hline
     $\widetilde{\chi}_{3}^{0}$&$799$&$\widetilde{\nu}_{e/\mu}$&$498$&$\widetilde{\tau}_{1}$&$94$&$\widetilde{b}_{1}$&$823$&$\widetilde{d}_{R}$&$968$&$m_{H^{\pm}}$&$872$\\ \hline
    $\widetilde{\chi}_{4}^{0}$&$802$&$\widetilde{\nu}_{\tau}$&$486$&$\widetilde{\tau}_{2}$&$496$&$\widetilde{b}_{2}$&$927$&$\widetilde{d}_{L}$&$1013$&$\widetilde{g}$&$616$\\ \hline
		\end{tabular}
		\label{tab:masses_2}
\end{table}

\begin{table}[ht]
  \small \centering
	\caption{Spectrum (in GeV) for the $\tan \beta$ = 20.5 benchmark point. 
Here, $M_{1/2}$ = 460 GeV, $M_{\rm V}$ = 2600 GeV, $m_{\rm t}$ = 173.4 GeV, $M_{\rm Z}$ = 91.187 GeV, $\Omega_{\chi}$ = 0.1107, $\sigma_{SI} = 2.9 \times 10^{-10}$ pb,
and $\left\langle \sigma v \right\rangle_{\gamma\gamma} = 4.2 \times 10^{-28} ~cm^{3}/s$. The central prediction for the $p \!\rightarrow\! {(e\vert\mu)}^{\!+}\! \pi^0$
proton lifetime is around $4.2 \times 10^{34}$ years. The lightest neutralino is 99.8\% bino.}
		\begin{tabular}{|c|c||c|c||c|c||c|c||c|c||c|c|} \hline		
    $\widetilde{\chi}_{1}^{0}$&$89$&$\widetilde{\chi}_{1}^{\pm}$&$193$&$\widetilde{e}_{R}$&$175$&$\widetilde{\rm t}_{1}$&$489$&$\widetilde{u}_{R}$&$925$&$m_{h}$&$119.7$\\ \hline
    $\widetilde{\chi}_{2}^{0}$&$193$&$\widetilde{\chi}_{2}^{\pm}$&$787$&$\widetilde{e}_{L}$&$500$&$\widetilde{\rm t}_{2}$&$876$&$\widetilde{u}_{L}$&$1005$&$m_{A,H}$&$849$\\ \hline
     $\widetilde{\chi}_{3}^{0}$&$782$&$\widetilde{\nu}_{e/\mu}$&$494$&$\widetilde{\tau}_{1}$&$98$&$\widetilde{b}_{1}$&$822$&$\widetilde{d}_{R}$&$961$&$m_{H^{\pm}}$&$853$\\ \hline
    $\widetilde{\chi}_{4}^{0}$&$786$&$\widetilde{\nu}_{\tau}$&$482$&$\widetilde{\tau}_{2}$&$492$&$\widetilde{b}_{2}$&$920$&$\widetilde{d}_{L}$&$1008$&$\widetilde{g}$&$637$\\ \hline
		\end{tabular}
		\label{tab:masses_3}
\end{table}

\begin{table}[ht]
  \small \centering
	\caption{Spectrum (in GeV) for the $\tan \beta$ = 21.0 benchmark point. 
Here, $M_{1/2}$ = 555 GeV, $M_{\rm V}$ = 2025 GeV, $m_{\rm t}$ = 174.3 GeV, $M_{\rm Z}$ = 91.187 GeV, $\Omega_{\chi}$ = 0.1150, $\sigma_{SI} = 1.3 \times 10^{-10}$ pb,
and $\left\langle \sigma v \right\rangle_{\gamma\gamma} = 2.2 \times 10^{-28} ~cm^{3}/s$. The central prediction for the $p \!\rightarrow\! {(e\vert\mu)}^{\!+}\! \pi^0$
proton lifetime is around $4.5 \times 10^{34}$ years. The lightest neutralino is 99.9\% bino.}
		\begin{tabular}{|c|c||c|c||c|c||c|c||c|c||c|c|} \hline		
    $\widetilde{\chi}_{1}^{0}$&$108$&$\widetilde{\chi}_{1}^{\pm}$&$234$&$\widetilde{e}_{R}$&$209$&$\widetilde{\rm t}_{1}$&$603$&$\widetilde{u}_{R}$&$1114$&$m_{h}$&$121.6$\\ \hline
    $\widetilde{\chi}_{2}^{0}$&$234$&$\widetilde{\chi}_{2}^{\pm}$&$944$&$\widetilde{e}_{L}$&$603$&$\widetilde{\rm t}_{2}$&$1036$&$\widetilde{u}_{L}$&$1211$&$m_{A,H}$&$1021$\\ \hline
     $\widetilde{\chi}_{3}^{0}$&$940$&$\widetilde{\nu}_{e/\mu}$&$598$&$\widetilde{\tau}_{1}$&$117$&$\widetilde{b}_{1}$&$991$&$\widetilde{d}_{R}$&$1157$&$m_{H^{\pm}}$&$1025$\\ \hline
    $\widetilde{\chi}_{4}^{0}$&$943$&$\widetilde{\nu}_{\tau}$&$584$&$\widetilde{\tau}_{2}$&$592$&$\widetilde{b}_{2}$&$1104$&$\widetilde{d}_{L}$&$1213$&$\widetilde{g}$&$754$\\ \hline
		\end{tabular}
		\label{tab:masses_4}
\end{table}

%%%%%%%%%%%%%%%%%%%%%%%%%%%%%%%%%%%%%%%%%%%%%%%%%%%%%%%%%%%%%%%%

\section{A Counting of Parameters\label{sct:acount}}

We close this paper with a brief retrospective dissection of the strong correlations 
among the major parameters in our model, in an attempt to better
elucidate the physics underlying our numerical results. For simplicity, we will not 
consider the SM fermion masses except for the top quark. Thus, we have $\tan\beta$
and eight mass parameters in our model, with gravity mediated supersymmetry 
breaking: $M_{1/2}$, $M_0$, $A$, $B_{\mu}$, $\mu$, $M_{\rm V}$, the $Z$-boson mass $M_{\rm Z}$,
and $m_{\rm t}$. The No-Scale boundary condition gives $M_0=A=B_{\mu}=0$.
There are two degrees of freedom eliminated by the EWSB conditions, one each for
minimization with respect to the neutral up- and down-like Higgs components.
From this, we may establish that two parameters, say $\tan\beta$ and $M_{\rm Z}$,
are functions of those remaining, namely $M_{1/2}$, $\mu$, $M_{\rm V}$, and $m_{\rm t}$.
Stipulating that the observed (near) equivalence between $\mu$ and $M_{1/2}$
may have a deeper theoretical motivation, we may tentatively also remove $\mu$
from the list of free parameters.  If we then revert to experimental
values, modulo some small uncertainties, for $M_{\rm Z}$ and $m_{\rm t}$, the specification
of a numerical value of $M_{\rm Z}$ will provide a relationship
between $M_{1/2}$ and $M_{\rm V}$. This leaves only a single free
parameter, the mass $M_{1/2}$.

A curious, and somewhat counter-intuitive, consequence of the prior
is that the region satisfying either the bare-minimal constraints
(driven by application of the very constrained 7-year WMAP dark matter density limits)
or the Super No-Scale condition (establishing $M_{1/2}$ as a function of $\mu$, $M_{\rm V}$, and $m_{\rm t}$)
can be quite large, as depicted in Figure~\ref{fig:Wedge_BareMinimal}.
Note that $\mu \simeq M_{1/2}$, so we will not have a mass parameter after we fix
the experimental values with uncertainties for $M_{\rm Z}$ and $m_{\rm t}$,
since the $M_{\rm Z}$ equation determines $M_{\rm V}$. In particular, the vector-like 
particles only contribute to the model structure via the renormalization group
equations for the gauge couplings and gaugino masses, so small uncertainties
in other parameters, such as $m_{\rm t}$, will propagate into large uncertainties
in $M_{\rm V}$. Moreover, because the minimum minimorum is determined
from the one-loop effective Higgs potential, $M_{1/2}$ is sensitive to $M_{\rm Z}$
and $\tan\beta$ as well.

%%%%%%%%%%%%%%%%%%%%%%%%%%%%%%%%%%%%%%%%%%%%%%%%%%%%%%%%%%%%%%%%

\section{Conclusions}

We have revisited the construction of the viable parameter space of No-Scale
$\cal{F}$-$SU(5)$, employing an updated numerical algorithm to significantly enhance the
scope, detail and accuracy of our prior study.

By sequential application of a set of ``bare-minimal'' phenomenological constraints,
considered to be of such experimental stability or intrinsic
theoretical necessity that the trespass of a single criterion should invalidate
the corresponding parameterization, we have comprehensively mapped the phenomenologically plausible region.
Specifically, these constraints consist of compliance with
i) the dynamically established high-scale boundary conditions $M_0=A=B_{\mu}=0$ of No-Scale Supergravity,
ii) consistent radiative electroweak symmetry breaking,
iii) precision LEP constraints on the lightest CP-even Higgs boson $m_{h}$ and other light SUSY chargino and neutralino mass content,
iv) the world average top-quark mass $172.2~{\rm GeV} \leq m_{\rm t} \leq 174.4~{\rm GeV}$, and
v) the 7-year WMAP limits 0.1088 $\leq \Omega_{\rm CDM} \leq$ 0.1158 with a single, neutral LSP as the CDM candidate.

A second category of phenomenological constraints, considered to be somewhat more ductile,
are associated with limits on the SUSY contributions to key rare processes, including
the flavor-changing neutral current decays $b \to s\gamma$ and $B_{s}^{0} \to \mu^+\mu^-$, and
loops affecting the muon anomalous magnetic moment $(g-2)_\mu$.  The model subspace that
is compatible with these additional criteria has been referred to as the ``golden strip'' of
No-Scale $\cal{F}$-$SU(5)$, featuring ${\rm Br}( b \to s \gamma) \geq 2.86 \times 10^{-4}$
and $\Delta a_\mu \geq 11 \times 10^{-10}$.

As a check of broad compatibility between the bottom-up phenomenological perspective (as
specifically driven by the CDM relic density) and the top-down theoretical perspective,
we have selected a portion of the model space for further study under application
of the ``Super No-Scale'' condition.  Specifically, this procedure compares the minimum $V_{EW}^{\rm min}$ 
of the scalar Higgs potential (after consistent electroweak symmetry breaking is enforced)
along a continuously connected string of adjacent model parameterizations, selecting out 
the model with the smallest locally bound value of $V_{EW}^{\rm min}$.  This point of secondary
minimization is referred to as the ``minimum minimorum''.  For fixed $M_{\rm V}$
and $m_{\rm t}$, the value of $V_{EW}^{\rm min}$ may be compared among interconnected (singly parameterized)
triplets of the free parameters $M_{1/2}, \tan\beta,~{\rm and}~M_{\rm Z}$, dynamically selecting a
single such combination.  The resulting dynamic determination is indeed in excellent agreement with
phenomenologically based selections for the matching $M_{\rm V}$ and $m_{\rm t}$. 

With the implementation of a more precise numerical algorithm and the advent of more powerful and comprehensive scanning technology, coupled to a philosophical shift toward the sequentially minimal application of constraints, we have here simultaneously widened our scope and narrowed our focus.
Whereas we had previously been (justifiably) content to ascribe a $(6\text{--}7)$ GeV shift away from the
minimum minimorum (corresponding to an absolute shift in $\tan \beta$ of about $5$) to higher order effects,
systematic to flaws in either the procedure or our approximation of the underlying physics~\cite{Li:2010uu}, that disagreement between
theory and phenomenology ({\it cf.}~Figure~\ref{fig:Vmin_4plex}) has now been reduced to a level significantly less than 1~GeV,
approximately a twenty-fold improvement.  Although we must yet consider these variations to be within the boundaries of error on our basic predictive capacity, we cannot resist taking some cheer from the inescapable observation that the central values of the phenomenologically
and theoretically favored regions demonstrate a significantly enhanced precision of agreement.
We consider the shifts that we have been pressed here to adopt relative to our prior work,
including somewhat elevated values for $\tan\beta$ and $M_{\rm V}$ (although the latter remains
enforcibly proximal to the scale of electroweak physics to alleviate any hierarchy concerns)
to be more than offset in fair trade by the improved resolution of this convergence.

The lightest CP-even Higgs boson mass is consistently predicted to be about 120~GeV~\cite{Li:2011xg}, neglecting a possible
upward shift of 3--4~GeV which may be induced by radiative loops in the vector-like fields~\cite{Li:2011ab}.
The predominantly bino flavored lightest neutralino is suitable for direct detection by the Xenon collaboration.
The partial lifetime for proton decay in the leading ${(e|\mu)}^{+} \pi^0 $ channels is distinctively rapid,
possibly as low as $(3\text{--}4) \times 10^{34}$~Years, just outside the current bounds of detection, and certainly testable
at the future Hyper-Kamiokande~\cite{Nakamura:2003hk} and DUSEL~\cite{Raby:2008pd} facilities.
The characteristic No-Scale $\cal{F}$-$SU(5)$ mass hierarchy, featuring a light stop and gluino, both lighter than
all other squarks, is quite stable, and is responsible for a distinctive collider
signal of ultra-high multiplicity of hadronic jets which is testable at the early LHC~\cite{Li:2011hr,Maxin:2011hy}.
This spectral ordering is not precisely replicated by any of the ``Snowmass Points and Slopes'' (SPS) benchmarks~\cite{Allanach:2002nj},
suggesting that it may be a highly distinctive feature.  The dexterity with which No-Scale $\cal{F}$-$SU(5)$ surmounts its
phenomenological hurdles is made all the much more remarkable by comparison
to the standard mSUGRA based alternatives, which despite a significantly greater freedom of parameterization, are being
rapidly cut down by the early emerging results from the LHC.

%%%%%%%%%%%%%%%%%%%%%%%%%%%%%%%%%%%%%%%%%%%%%%%%%%%%%%%%%%%%%%%%%%%%%%%%%%%%

\begin{acknowledgments}
This research was supported in part 
by the DOE grant DE-FG03-95-Er-40917 (TL and DVN),
by the Natural Science Foundation of China 
under grant numbers 10821504 and 11075194 (TL),
by the Mitchell-Heep Chair in High Energy Physics (JAM),
and by the Sam Houston State University
2011 Enhancement Research Grant program (JWW).
We also thank Sam Houston State University
for providing high performance computing resources.
\end{acknowledgments}

%%%%%%%%%%%%%%%%%%%%%%%%%%%%%%%%%%%%%%%%%%%%%%%%%%%%%%%%%%%%%%%%%%%%%%%%%%%%

\bibliography{bibliography}

\end{document}